\documentclass[twocolumn]{aastex63}

\newcommand{\xgen}{\textit{XGEN}}
\newcommand{\coup}{\textit{COUP}}
\newcommand{\fancyN}{$\mathcal{N}$}
\newcommand{\fancyf}{$\mathcal{F}$}
\newcommand{\water}{H$_2$O}
\newcommand{\hh}{H$_2$}

\newcommand{\htp}{H$_3^+$}
\newcommand{\change}{$\mathcal{C}$}

\shorttitle{X-ray Flare Driven Disk Chemistry}
\shortauthors{Waggoner \& Cleeves}

\usepackage{lineno}
\usepackage[caption=false]{subfig}

\begin{document}

\title{Classification of X-ray Flare Driven Chemical Variability in Protoplanetary Disks}

\correspondingauthor{Abygail R. Waggoner}
\email{arw6qz@virginia.edu}

\author[0000-0002-1566-389X]{Abygail R. Waggoner}
\affiliation{University of Virginia, Department of Chemistry, Charlottesville, VA 22904, USA}
\author[0000-0003-2076-8001]{L. Ilsedore Cleeves}
\affiliation{University of Virginia, Department of Astronomy, Charlottesville, VA 22904, USA}
\affiliation{University of Virginia, Department of Chemistry, Charlottesville, VA 22904, USA}

\begin{abstract}

Young stars are highly variable in the X-ray regime. In particular, bright X-ray flares can substantially enhance ionization in the surrounding protoplanetary disk. Since disk chemical evolution is impacted by ionization, X-ray flares have the potential to fundamentally alter the chemistry of planet forming regions. 
We present 2D disk chemical models that incorporate a stochastic X-ray flaring module, named \xgen, and examine the flares' overall chemical impact compared to models that assume a constant X-ray flux. 
We examine the impact of 500 years of flaring events and find global chemical changes
on both short time scales (days) in response to discrete flaring events and long time-scales (centuries) in response to the cumulative impact of many flares. 
Individual X-ray flares most strongly affect small gas-phase cations, where a single flare can temporarily enhance the abundance of species such as H$_3^+$, HCO$^+$, CH$_3^+$, and C$^+$. 
We find that flares { can also drive chemistry out of ``steady state'' over longer time periods}, 
where the disk-integrated abundance of some species, 
such as O and O$_2$, { changes by a few percent} over the 500 year model. We also explore whether the specific history of X-ray flaring events (randomly drawn but from the same energy distribution) impacts the chemical evolution and find that it does not. Finally, we examine the impact of X-ray flares on the electron fraction. While most molecules modeled are not highly sensitive to flares, certain species, including observable molecules, are very reactive to the dynamic environment of a young star.

\end{abstract}


\section{Introduction} 
\label{sec:intro}

The young central star in a protoplanetary disk plays an important role in shaping the disk's physical and chemical evolution. Solar mass pre-main sequence stars, i.e., T Tauri stars, are X-ray bright due to emission from hot ionized gas trapped in magnetic loops on the stellar surface \citep[e.g.,][]{feigelson1999,favata2005,feigelson2007}. 
The typical X-ray luminosity of a T Tauri star is $L_{\rm XR} \approx 2\times10^{30}$ erg s$^{-1}$ \citep{flaccomio2003b,wolk2005}, about $10^4$ times brighter than the modern day Sun \citep{peres2000}. 
In addition to being generally X-ray bright, 
T Tauri stars are also highly variable in X-rays. This variability arises at least in part due to X-ray flaring produced by magnetic reconnection events{ \citep[e.g.,][]{getman2008a,getman2008b}}. {The most powerful flares rapidly increase the star's X-ray luminosity by several orders of magnitude within a few hours, followed by exponentially decay over the course of days \citep[e.g.,][]{getman2021b}}. 

As a star ages, the stellar dynamo stabilizes and magnetic fields are thought to weaken, thus X-ray luminosity and variability becomes less intense and frequent
\citep[see review of][and references therein]{guedel2004}. 
Therefore, to best understand {how high energy processes impact planet formation}, we must understand how the highly variable X-ray radiation field of T Tauri stars affects the protoplanetary disk environment.

{In general,} high energy ionization, including from X-rays, drives disk chemistry at low ($<100$~K) temperatures through ion-neutral reactions \citep{strom1989,glassgold1997,haisch2001,fedele2010,cleeves2014par}.
X-ray ionization occurs primarily via the ionization of \hh\ and He \citep{maloney1996}. Specifically, the ionization of \hh\ results in the formation of H and \htp, both of which are essential in the formation of more complex molecules in both the gas and ice phases \citep[e.g.,][]{mccall2006}.
X-rays also have the potential to drive chemistry close to the disk mid-plane, since high energy ($\gtrsim5$~keV) X-ray photons are capable of penetrating denser gas { \citep[$N_H = 9 \times 10^{23}$ cm$^{-2}$;][]{bethell2011x}}. 
Less energetic photons, like UV, are absorbed closer to the disk surface \citep{glassgold2005,bethell2011u}.

Astrochemical codes typically include X-ray related processes often as a single value or as a single spectrum constant in time that are
representative of a ``characteristic'' ionization rate.
However, there is increasing theoretical and observational evidence that this simplification may not always be valid. {\citet{Ilgner2006flares} were first to model how periodic flares impact the ionization fraction of a disk both using a simplified network and a full chemical network. They found the size of a disk's ``dead zone'' was impacted by flares, but the extent of the variation was sensitive to the distribution and size of dust grains.}

{More recently, potential observational evidence of flares was reported in}  \citet{cleeves2017}. Specifically, significant (20 $\sigma$) changes in the emission of a known X-ray sensitive molecule, H$^{13}$CO$^+$, was discovered in the IM Lup protoplanetary disk. With simple chemical models, they found that X-ray flares were a viable explanation for this variability. {Since H$^{13}$CO$^+$ is not the only X-ray sensitive molecule,}  \citet{waggoner2019} carried out a theoretical study and found that gas-phase \water\ can also experience a short-lived enhancement in abundance in response to a single, strong flaring event that enhances the X-ray luminosity by a factor of 100. Chemical evolution in planet-forming disks driven by X-ray flares has otherwise been relatively unexplored. 

We present a comprehensive study on 2D disk chemical responses (radius and height) to 500 years of stochastic X-ray flaring events. Section \ref{sec:xgen} introduces a new, flexible X-ray flare model, \xgen, which is used to model a stochastic X-ray light curve for a T Tauri star. The light curve is incorporated into a chemical disk model described in Section \ref{sec:diskmodel}.
The results from the model are presented in Section \ref{sec_results}, where chemical responses are categorized and relevant reactions discussed in depth. 
In Section \ref{sec_disc}, we discuss the results, such as the short- and long-term impact of flares on disk chemistry and electron abundances, along with their observational implications. Lastly, in Section \ref{sec_conc} a summary of the results and concluding remarks are provided.

\section{Model}\label{sec_model}

\subsection{X-ray Light Curve Generator: XGEN}\label{sec:xgen}

We present a flexible code for generating randomized light curves drawing from known stellar X-ray flare statistics. The code, called the \textit{X-ray Light Curve Generator} (\xgen)\footnote{\xgen\ is available upon request from A.R.W.}, models a stochastic light curve based on a user-provided flare frequency, flare energy distribution, and rise/decay time. Literature has shown that the energy distribution of stellar flares in an X-ray light curve can be defined by a power-law distribution:
\begin{equation}
    {dN \over dE} = \beta log(E_{\rm tot})^{1.0-\alpha},  \label{eq_powerlaw}
\end{equation}
where the power-law defines the total number of flares ($dN$) that occur over an energy range $dE$ \citep[e.g.,][]{Hudson1991,caramazza2007}. $E_{\rm tot}$ signifies the total energy output of a single flare. $\beta$ is a normalization factor that controls the total number of flares the model produces. As described below, the time evolution of individual flares is represented by a sharp exponential rise followed by a slower exponential decay. While the present paper applies the code to the case of T Tauri-like stellar parameters, we note that \xgen\ is not limited to any particular type of star. Instead, \xgen\ can create any light curve based on a power law energy distribution, including but not limited to flares from main sequence stars or active galactic nuclei.

\subsubsection{Overview of Model}

\xgen\ uses a random number generator to determine the probability that a flare with total energy $E_{\rm tot}$ will occur. $E_{\rm tot}$ is spectrally integrated from 1 - 20 keV{. The spectrum used in this work is considered a typical X-ray emission spectrum for a T-Tauri star, as was observed by \textit{Swift} in the IM Lup protoplanetary disk \citep[][Figure 2]{cleeves2017}.} {\xgen\ simulates flares by uniformly {scaling} the X-ray spectrum based on the flare magnitude. In reality, flares change the X-ray spectrum by producing proportionally more high-energy, or `hard,' photons relative to low-energy, or `soft,' photons. This work uses a simplified model, where hard and soft photons are equivalently increased, and the spectral energy distribution remains constant. X-ray hardening will be incorporated into future versions of \xgen.}

$E_{\rm tot}$ is discretized into energy ``bins'' within a specific range, so that a finite number of flare energies are considered possible. The lower and upper boundaries for each energy bin $E_{\rm tot}$ are defined as $E_{\rm low}$ and $E_{\rm up}$, respectively. The spread of $E_{\rm tot}$ per energy bin is defined as $\Delta E$, where $\Delta E = E_{\rm up} - E_{\rm low}$. In this work, the energy bins are spaced evenly with $\Delta E=10^{0.01}$ erg. {Note that $E_{\rm low}$ and $E_{\rm up}$ span the total energy range considered flares, from $E_{\rm min}$ to $E_{\rm max}$ and all energy values in between at a resolution of $\Delta E$.}
The random number generator uses the python 2.7 numpy.random package, version 1.16.6 and build py27hbc911f0\_0. 

The probability of a flare with energy $E_{\rm tot}$ occurring is defined by
\begin{equation}\label{eq_prob}
    P(E_{\rm tot}) = \beta \, (E_{\rm up}^{-\alpha+1} - E_{\rm low}^{-\alpha+1})
\end{equation}
\begin{equation}
    \beta = -\mathcal{F}\; E_{min}^{\alpha-1}\label{eq_shape} \; \Delta t 
\end{equation}
where $\beta$ is a normalization constant, $\Delta t$ is the time step resolution ran in \xgen, and \fancyf\ is the target flare frequency.
Multiple flares are allowed to occur within the same time step, which represents flares occurring simultaneously on different parts of the star. 
Model parameters and their symbols are provided in Table \ref{table_xgen_parameters}.

All flares are assumed to have an exponential rise ($\tau_{\rm rise}$) and decay ($\tau_{\rm decay}$) time profile. \xgen\ is written such that $\tau_{\rm rise}$ and $\tau_{\rm decay}$ can be variable based on a probability distribution, but for simplicity, we assume a uniform $\tau_{\rm rise}=3$\,hr and $\tau_{\rm decay}=8$\,hr for all flares. Given these timescales and a total flare energy, $E_{\rm tot}$, one can use the following to solve for useful parameters such as the peak change in luminosity, $\Delta L_{\rm peak}$, and the X-ray luminosity at a given time:  
\begin{equation}
    E_{tot} = \Delta L_{\rm peak}\int^{t_{peak}}_{-\infty}e^{t/\tau_{\rm rise}}dt 
    + \Delta L_{\rm peak}\int_{t_{peak}}^{\infty}e^{-t/\tau_{\rm decay}}dt 
    \label{eq_flare_shape}
\end{equation}

The cumulative light curve is constructed by adding the luminosity of all individual flares at every time step, then adding in a ``baseline'' or characteristic luminosity ($L_{\rm char}$). For the purpose of this study, light curves are also normalized with respect to $L_{\rm char}$ and represented by $\Delta L_{\rm XR}$.
\begin{equation}
    \Delta L_{\rm XR} = {{L_{\rm char} + \sum L_{\rm flare}} \over L_{\rm char}}
    \label{eq_lxr}
\end{equation}

\begin{figure}
    \centering
    \includegraphics[scale=0.56]{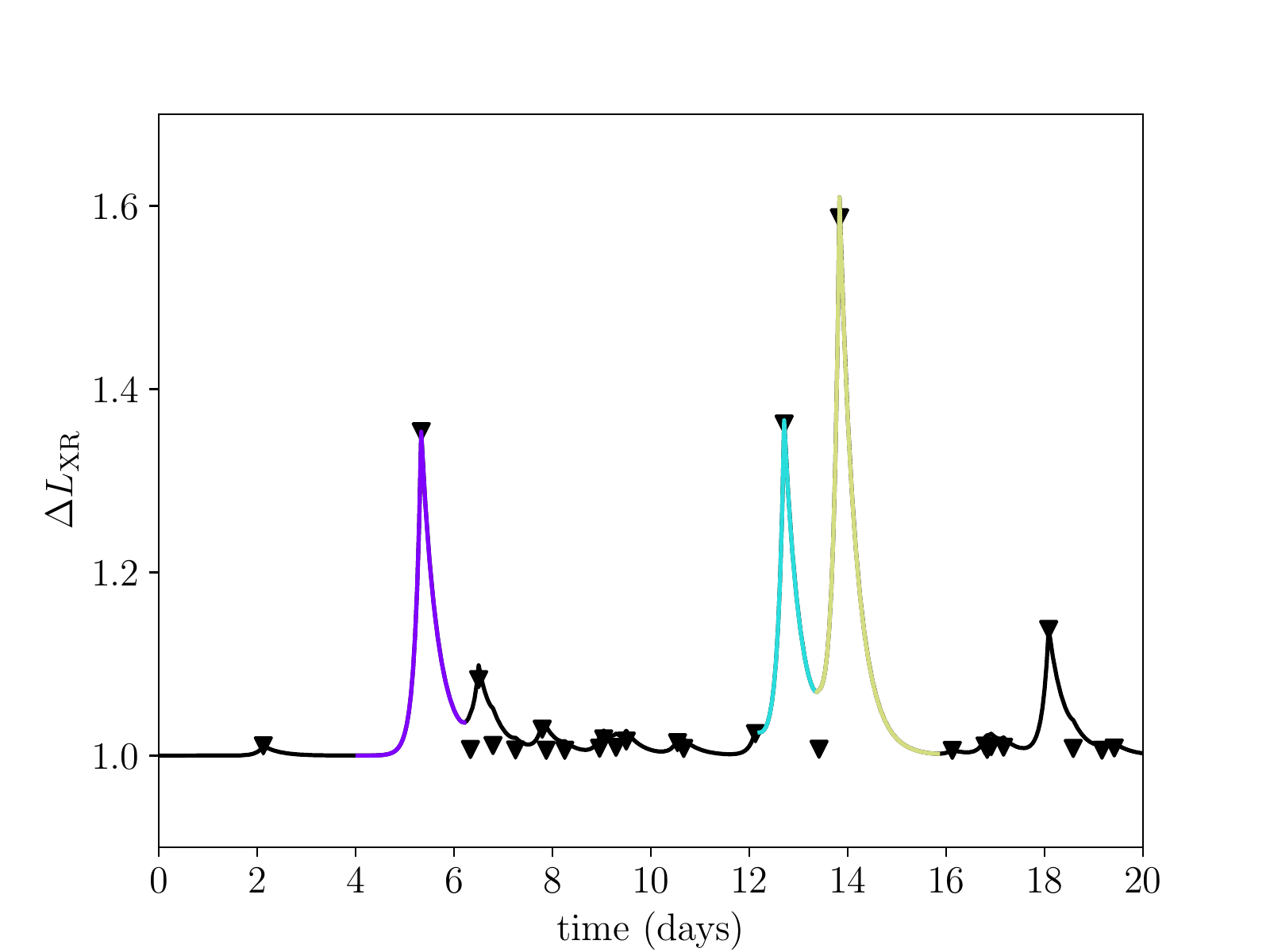}
    \caption{A simulated light curve for a T Tauri star. Individual modeled flare peaks, including unobservable flares, are represented by triangles. Individual observable flares (distinguishable peak and $E_{tot}>10^{34}$ erg) are highlighted as purple, blue, and yellow. Note there are 27 modeled flares but only 3 observable flares based on our criteria.}
    \label{fig:observable_flares} 
\end{figure}

\xgen\ includes flares that are lower than presently detectable, e.g., microflaring and nanoflaring events. Therefore many of the individually modeled flares overlap or are below a realistic detection limit. As a result, we define `observable flares' as those that satisfy two criteria. First, the flare peak must be distinguishable. \xgen\ identifies individual flare peaks as any point were the slope is effectively zero
(for practical purposes, where $|dL_{\rm XR}/dt| < 0.015$). Once individual flare peaks are identified, the beginning and end of each flare is determined by a location where the slope either returns to zero or switches signs. This process constructs each distinguishable flare. Second, the distinguishable flare must have a total energy greater than a predetermined minimum energy value ($E_{\rm min,obs}$) to be considered observable (Figure \ref{fig:observable_flares}). 
\xgen\ can then determine energy distribution and frequency of observable and distinguishable flares, which can then be compared to an observed statistical distribution (Figure \ref{fig_edist}). 

\subsubsection{X-ray Light Curve for a T Tauri Star}\label{sec:xgen_ttauri}

The light curves presented in this work are modeled after the statistical analysis of solar type young stars presented in \citet{wolk2005} from the \textit{Chandra Orion Ultra-deep Project} (\coup). The \coup\ survey is the longest, continuous observation of stars in the X-ray regime, thus making it the most comprehensive study of X-ray flaring events in solar mass stars to date \citep{getman2005}. 
The energy distribution of flares observed by 
\citet{wolk2005} are best fit by
\begin{equation}
    N = 1.1 \, {\rm log}(E_{\rm tot}) ^ {-0.66} \label{eq_wolk_edist}
\end{equation}
where $N$ is the cumulative number of flares observed with total energy $E_{\rm tot}$ or greater and, when re-cast in $dN/dE$, representative of $\alpha=1.66$. 

The observed average flare frequency by \citet{wolk2005} is approximately 1 flare every 650\,ks, with an uncertainty of 10$\%$, or $\sim$ 50 flares per year.
The modeled light curves are in agreement with an average of \fancyf\ $\sim 64 {\rm yr}^{-1}$. This is a slightly higher flare frequency than the target frequency, {but it is only natural that the modelled flare count is higher than the observed flare count. Since the modelled flare count comes from a direct comparison to the observed data, the excess flares are attributed to smaller low energy flares that would not be distinguishable due to observational limitations.}

\begin{deluxetable}{ccc}
\tablecolumns{3}
\tablewidth{0pt}
\tablecaption{Input parameters used to generate an X-ray flare light curve for a T Tauri star. 
    \label{table_xgen_parameters}}
\tablehead{
\colhead{Quantity} 
& \colhead{Symbol} & \colhead{Value}}
\startdata
Max Flare Energy$^*$ & $E_{\rm max}$            &   $10^{37.57}$\,erg      \\
Min. Flare Energy$^*$ & $E_{\rm min,obs}$    &   $10^{34.0}$\,erg             \\
Target Flare Frequency$^*$ &  \fancyf     &   48.5 yr$^{-1}$    \\
Characteristic Luminosity & L$_{\rm char}$   &  $10^{30.25}$\,erg{s$^{-1}$}      \\
Flare rise time & $\tau_{\rm rise}$    &   $3$\,hrs             \\
 Flare decay time & $\tau_{\rm decay}$  &   $8$\,hrs             \\
\hline
\multicolumn{3}{c}{Parameters quantitatively fit$^{**}$} \\
\hline
Power-law index & $\alpha$ &   1.64 \\
Min. modeled flare energy & $E_{\rm min,model}$ &   $10^{32.50}$\,erg \\
Energy Step Resolution & $\Delta E$  &   $10^{0.01}$ erg
\enddata
\tablecomments{* parameters based on a statistical analysis of young solar mass stars presented in \cite{wolk2005}. ** parameters found to produce the best fit energy distribution (Figure \ref{fig_edist}).}
\end{deluxetable}

Flares modeled by \xgen\ range from $E_{\rm min}=10^{32.50}$\,erg to $E_{\rm max}=10^{37.57}$\,erg. 
{Flares with $E_{\rm tot} < 10^{32.50}$\,erg corresponded to $\Delta L_{\rm XR}$ values less than 1.004, and are considered negligible for the purpose of this model. While $10^{32.50}$\,erg also corresponds to a negligible flare peak ($\Delta L_{\rm XR} \sim 1.005$), we found that larger $E_{\rm min}$ values were unable to fit the desired energy distribution as well.}
The minimum total energy for an observable flare was set to $E_{\rm tot}$ of the weakest flare reported in \citet{wolk2005}, 
$E_{\rm min,obs}=10^{34.0}$\,erg.
The maximum allowed flare energy in the model is consistent with the largest flare energy detected in \citet{wolk2005} 
($E_{tot}=10^{37.57}$\,erg){, and is in agreement with the `super-flares' detected in \citet{getman2021a}.} 
The characteristic luminosity is set to $L_{\rm char}=10^{30.25}$\,erg{s$^{-1}$} \citep[e.g.,][]{flaccomio2003b}. {The maximum flare energy considered is thus equal to the amount of energy output by the star at quiescence for a period of 8 months.}

\begin{figure}
    \centering
    \includegraphics[scale=0.57]{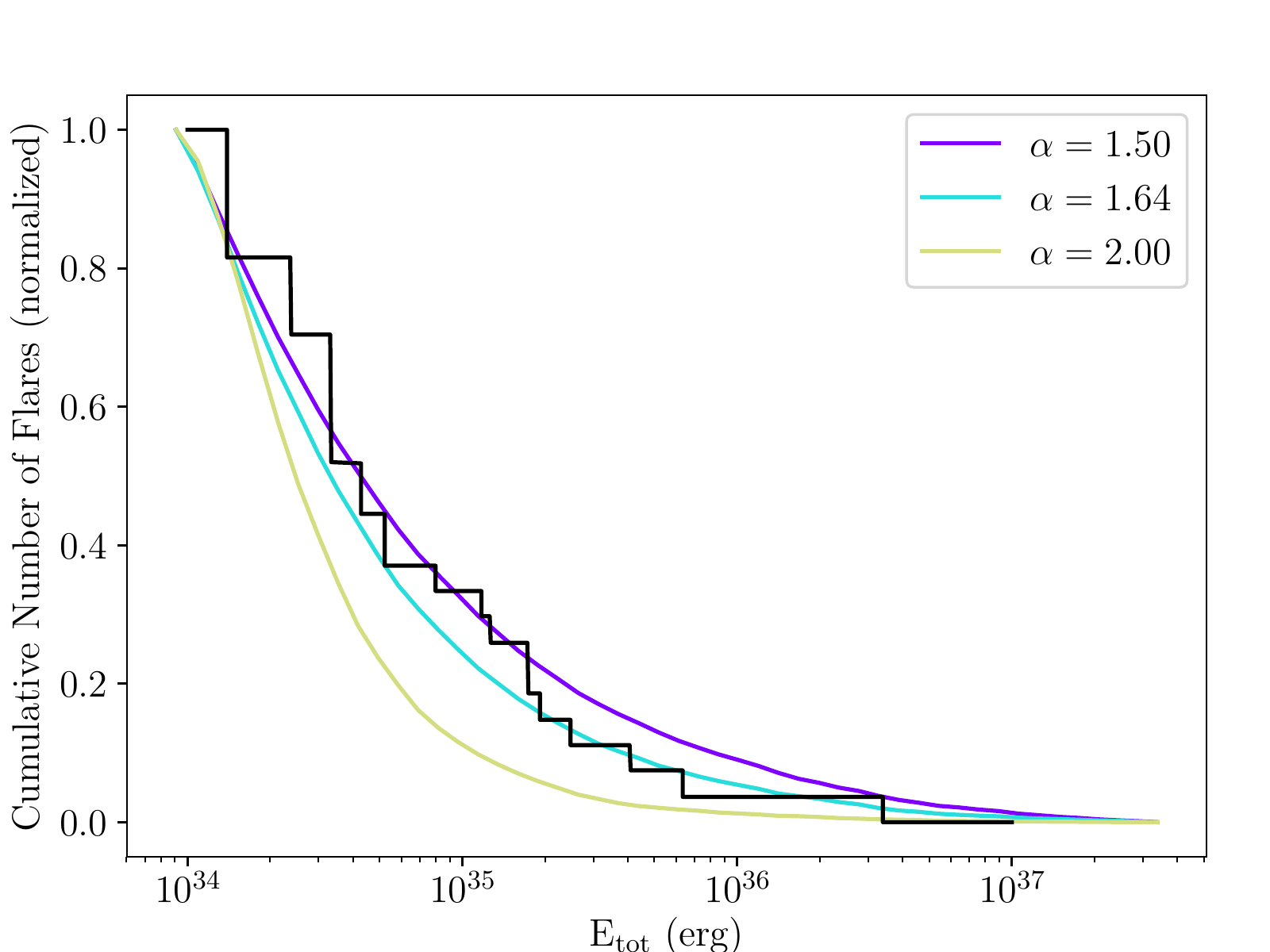}
    \caption{
    Energy distribution of flares from a T Tauri star compared to the observations of \citet{wolk2005} shown in black (see their Figure 9). 
    Over-plotted are the simulated flare energy distributions for $\alpha=1.50$, $1.64$, and $2.00$ tabulated for a 100 year light curve. A power law index of $\alpha = 1.64$ was found to yield the best match to the observations.}
    \label{fig_edist}
\end{figure}

To best match the \citet{wolk2005} distribution, a series of different power laws were tested with $\alpha$ values ranging from 1.5 to 3.0. For each value of $\alpha$, 100 one-year curves were produced and analyzed at one hour time step resolution.
The energy distribution of observable flares is sensitive to the chosen $\alpha$ value, as shown in Figure \ref{fig_edist}. 
We find that $\alpha$ = 1.64 yielded the best overall fit to the observed energy distribution of solar mass YSOs with $\chi^2=0.15$, where $\chi^2$ measures the goodness of fit.
This power law index is consistent with the best fit of $\alpha = 1.66$ reported in \citet[][Equation \ref{eq_wolk_edist}]{wolk2005}.
It should be noted that $\alpha =1.50$ appears to better fit the observed distribution below $E_{\rm tot} \sim 10^{34.5}$\,erg. However, $\alpha$ values less than 1.64 fail to fit the high energy flares and have higher $\chi^2$ values. For example, $\chi^2 = 0.91$ for $\alpha=1.50$.
{Refitting the light curve to the best fit 
$\alpha$
value, rather than using 
$\alpha = 1.66$, 
was necessary since \citet{wolk2005} empirically derived alpha from data. However, the synthetic light curves in this work include microflaring and flare blending, neither of which can be identified in observations.}

\begin{figure*}
    \centering
    \includegraphics[scale=0.78]{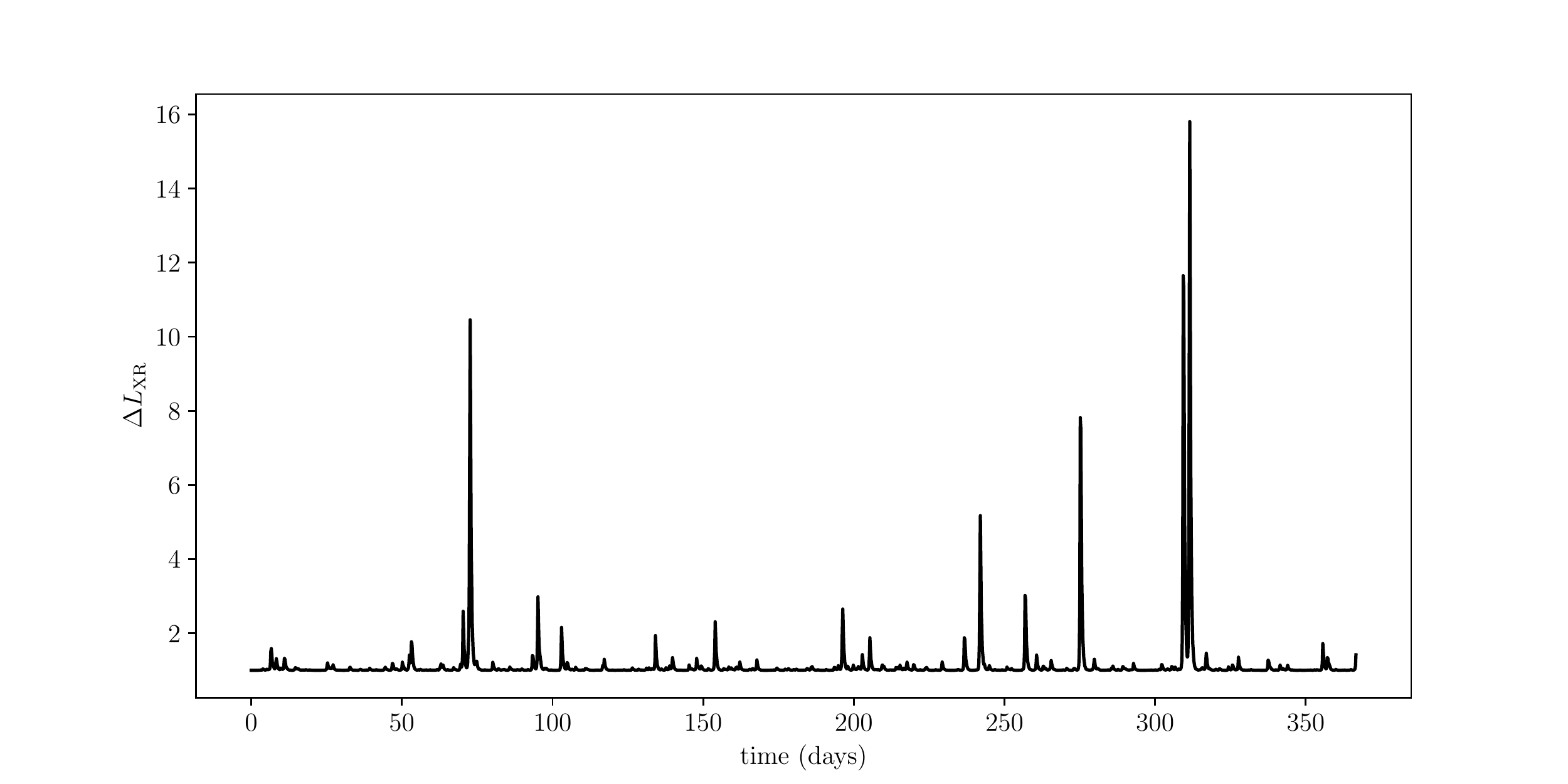}
    \caption{Year one of a 500 year X-ray light curve
    produced by \xgen\ for T Tauri-like stellar parameters.
    $\Delta L_{\rm XR}$ is the relative (multiplicative) change in luminosity when there is a flare compared to the characteristic luminosity (Equation \ref{eq_lxr}).}
    \label{fig_curve}
\end{figure*}

Incorporating all of the effects described above and adopting the observationally motivated X-ray flaring statistics, Figure~\ref{fig_curve} shows a one year subsection of our model light curve. As can be seen, moderate flares (changes greater than 10 times $L_{\rm char}$) happen a few times (in this case 3 times) over the course of a year, while weaker flares (changes of 2 to 3 times $L_{\rm char}$) are far more frequent.

\subsection{Disk Chemical Model}\label{sec:diskmodel}

The chemical disk model used in this work is modeled after \citet{fogel2011}, which adopts the rate equation method and includes 644 chemical species and 5944 types of chemical and physical processes. 
The physical environment is modeled after the IM Lup protoplanetary disk system, the original source observed to experience variability \citep{cleeves2017}. The disk and stellar physical parameters are representative of a solar mass star between $0.5 - 1$ Myr old \citep{cleeves2016}.

A two dimensional disk is simulated by running a series of point locations at various radial distances ($R$) and vertical heights ($Z$).
We model the chemistry at radii of $R = 1$, $5$, $10$, $20$, $30$, $40$, $60$, $80,$ and $100$\,au from the central star. {We note that the gas in IM Lup extends to $R\sim1000$\,au, but we only model the inner disk because we find that chemical responses to flares beyond $100$\,au are generally negligible.} Modeled vertical heights (in cylindrical coordinates) range from the disk mid-plane to the disk surface with vertical height ratios of $Z/R = 0.0$, $0.1$, $0.2$, $0.3$, $0.4$, $0.5,$ and $0.6$ at each radii.

\begin{table}[]
    \centering
     \caption{Initial chemical abundances with respect to H for volatile species. These values are representative of molecular cloud abundances \citep{aikawa99}.}
    \label{tab:initial_abund}
    \begin{tabular}{lc|lc|lc}
         H$_2$      & $5.0\times 10^{-1}$    & C         & $5.0\times 10^{-9}$    & CS        & $4.0\times 10^{-9}$ \\
         H$_3^+$    & $1.0\times 10^{-8}$    & C$^+$     & $1.0\times 10^{-9}$    & SO        & $5.0\times 10^{-9}$ \\
         He         & $1.4\times 10^{-1}$    & C$_2$H    & $8.0\times 10^{-9}$    & S$^+$     & $1.0\times 10^{-11}$ \\
         O          & $1.0\times 10^{-8}$    & N$_2$     & $3.51\times 10^{-5}$   & Si$^+$    & $1.0\times 10^{-11}$ \\
         O$_2$      & $1.0\times 10^{-8}$    & NH$_3$    & $8.0\times 10^{-8}$    & Fe$^+$    & $1.0\times 10^{-11}$ \\
         CO         & $1.2\times 10^{-4}$    & HCN       & $1.0\times 10^{-8}$    &   & \\
         HCO$^+$    & $9.0\times 10^{-9}$    & CN        & $6.0\times 10^{-8}$    &  &    \\
         H$_2$O$_{\rm ice}$    & $8.0\times 10^{-5}$    &   Mg$^+$          &     $1.0\times 10^{-11}$      &          \\
    \end{tabular}
  
\end{table}

The model begins with chemical abundances considered typical for a molecular cloud \citep[motivated by ][see Table \ref{tab:initial_abund}]{aikawa99}. We run the model for $0.5$ Myr to achieve a pseudo steady state, then the X-ray flaring events produced by \xgen\ are initiated. 
Once flaring begins, defined as $t=0.0$ years in the Figures below, the model takes linear, four hour time steps {for 500 years}. 
Small time steps were necessary to ensure individual flares in the light curve were not `missed' in the model. We tested a range of time steps from $1$ minute to $2$ days, and found that four hours allowed us to obtain sufficiently long duration models without losing time resolution. We note that four hours means that we only have 2-3 time steps crossing the flare{;} however{,} based on our tests of shorter time steps, the chemical results were not significantly impacted by this. The implication of these tests is that the total flare energy is more important than the details of the shape of an individual flare.
{We found that a 500 year model was long enough to probe the diverse range of modeled flare energies while balancing the computational limits of the model. A 500 year model was long enough to ensure that at least one of the strongest possible flares occurred ($\Delta L_{\rm XR}>800$), and the chemical responses to such a flare, could be modeled. In our fiducial model, two such flares occurred.}

\begin{figure}
    \centering
    \includegraphics[scale=0.55]{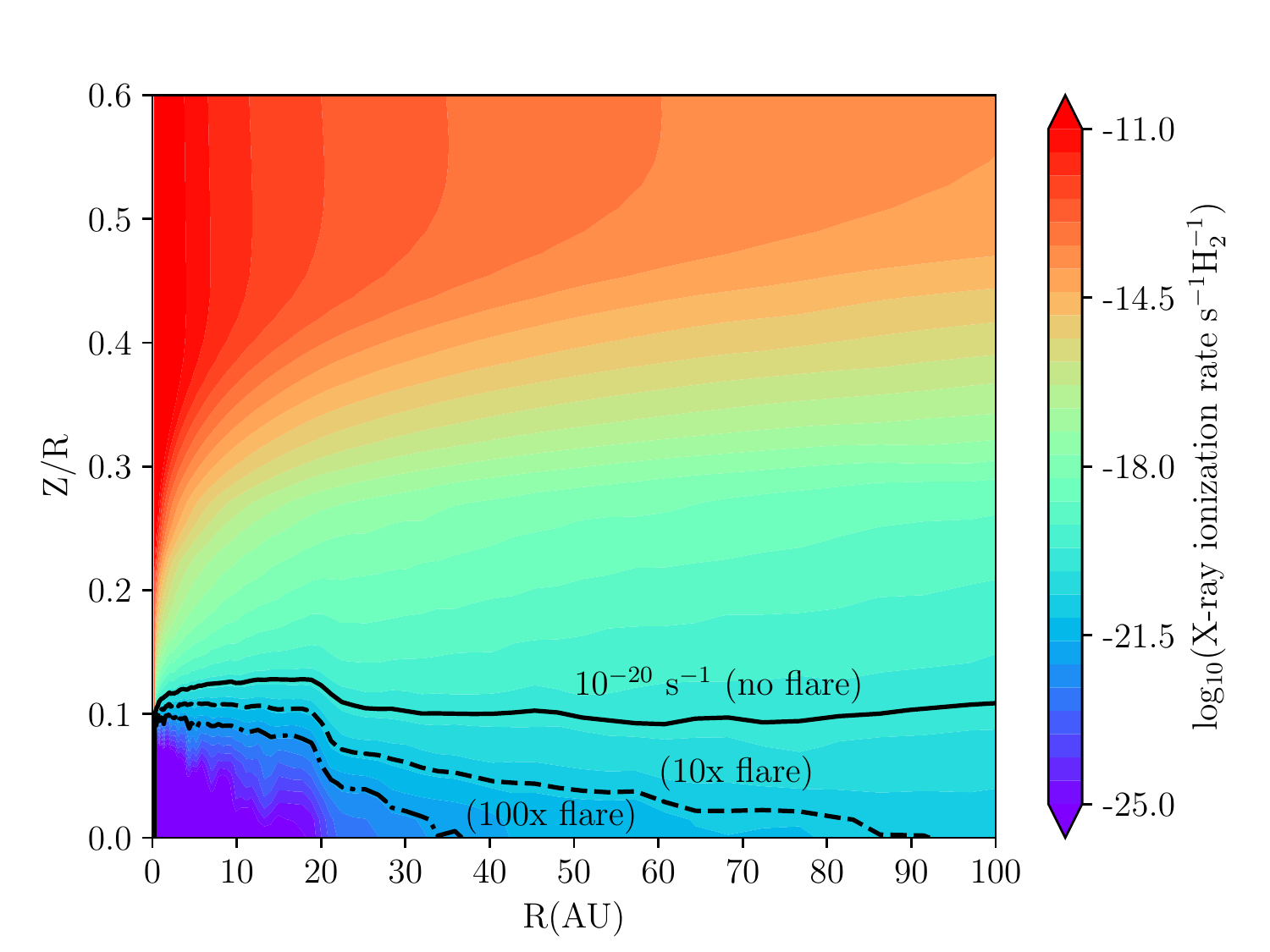}
    \caption{{Comparison of the X-ray ionization rate to the CR ionization rate. The background color contours show the non-flaring X-ray ionization rate in the disk. Overlaid are lines indicating where the X-ray ionization rate is {$10^{-20}$s$^{-1}$} for cases where $\Delta$L$_{\rm XR}$= 1, 10, 100, and 1000. Outside of 10~au, the CRs are unattenuated, and therefore these values can be directly compared to an assumed CR ionization rate (here $2\times10^{-20}$s$^{-1}$). Depending on flaring state, X-rays dominate above the contours and CRs below.
    }}
    \label{fig:xrate}
\end{figure}

Disk density, temperature, UV ionization, and X-ray ionization are extracted similarly as was done in \citet{waggoner2019} and are based off of the \citet{cleeves2016} physical structure. To summarize, the energy-dependent X-ray (and UV) radiative transfer is computed using \citet{bethell2011u}. The spectra are used to compute a local spectrally integrated UV flux, and for the X-rays, to compute a local H$_2$ and He ionization rate at every ($R,Z$) position in the disk. The model uses an incident cosmic ray ionization rate of $\zeta = 2\times10^{-20}$ s$^{-1}$ {\citep{cleeves2015}}. We note this value is lower than typical ionization rates of $\zeta \approx 10^{-17}$ s$^{-1}$; however, there is increasing evidence of low cosmic ray ionization rates within molecular gas disks {\citep[e.g.][]{seifert2021}}. 
{Figure \ref{fig:xrate} demonstrates regions of the disk where the X-ray ionization rate dominates the cosmic ray ionization rate.}

X-ray flares are incorporated into the chemical disk network by uniformly increasing the X-ray ionization rate{, and by extension the X-ray {flux},} by $\Delta L_{XR}$  during flaring events. {We emphasize that this assumption does not incorporate X-ray hardening, which will be explored in future work.}
{The chemical disk model assumes that flares do not have a directional preference, i.e., a flare propagates uniformly throughout the disk and is only attenuated. This assumption is valid if the X-ray illumination is both azimuthally symmetric and symmetric about the mid-plane.} This simplification is valid to first order because the size of X-ray emitting coronal loops on T Tauri stars is expected to be large compared to the size of the star, allowing the disk to be more uniformly illuminated. Directional flares will be explored in future work.  
The observed increase in H$^{13}$CO$^+$ in IM Lup supports this theory, as enhancement was the same in both blue- and red-shifted gas \citep{cleeves2017}.

{In this model X-rays interact with the disk by ionizing H$_2$ and He. While previous work has found that additional noble gasses, such as Ne and Ar, and metals have been seen to be sensitive to flares, we do not include these species in our network. 
However, there is no evidence that these species contribute to the formation of molecules \citep[e.g., ][]{adamkovis2011}, and they are not considered a main source of electrons in the disk when compared to, e.g., ionization of H$_2$ and He, as well as carbon ionization by UV. Therefore, omitting these species will not strongly impact our results. {We do note that the metal abundances used in this model ($10^{-11}$ w.r.t. H$_{\rm tot}$) are considered to be a low metal abundance compared to a typical metal abundance in disks \citep[e.g.,][]{fogel2011,aikawa99}. In future work we will explore increasing the abundances of these species, which can act as sources of electrons especially in regions of the disk where there is little atomic carbon, closer to the midplane.}}

\subsection{Assessment of Chemical Variability}

To best understand the impact of flares on chemistry in the disk, the column density, $N(R)$, and disk-integrated number, designated by \fancyN, of each species was calculated at one day resolution.
The integrated number of each species is found by assuming azimuthal symmetry, i.e., \fancyN$ = 2\pi \int R N(R) dR$. 
{Since it takes time for light to propagate from the central star to the disk, 
the affects of a flare do not occur instantaneously at all radial distances. 
For example, when a flare occurs it takes 8.32 minutes to reach $R=1$\,au 
and 13.86 hours to reach $R=100$\,au. To accurately model the chemical impact of flares and simulate light propagation through the disk, b}oth 
$N(R)$ and \fancyN\ incorporate light travel time as the flare propagates radially outward.
As a reference, the disk integrated number of hydrogen atoms (in H and H$_2$) is \fancyN$_{\rm hydrogen} = 9.46 \times 10^{55}$. Species with \fancyN$<10^{25}$ (approximately a fractional abundance $10^{-30}$ w.r.t. H{$_{\rm tot}$}) are considered below the numerical error of the model are are omitted from the following analysis. 

Species' level of susceptibility to individual flares is quantified by the standard deviation, $\sigma$, of the relative change in \fancyN, $\Delta$\fancyN, over the 500 year simulation time.
A large standard deviation is indicative of a highly variable species, and a small ($\approx 0$) standard deviation indicates that species is not significantly impacted by individual flares. 
The cumulative impact of many flares, i.e., the long-term (centuries) impact, is quantified by a relative change {in the total number of a particular species}, \change, at the end of the chemical model (${\mathcal{N}_{\rm final}}$) compared to the start (${\mathcal{N}_{0}}$) 
and is defined as: 
\begin{equation}
    \mathcal{C} = { {\mathcal{N}_{\rm final}} \over {\mathcal{N}_{0}}}.
\end{equation}
${\mathcal{N}_{\rm final}}$ and ${\mathcal{N}_{0}}$ are averaged from the first and final 20 time steps (3.3 days) of the model to ensure flares at the end or beginning of the model do not cause an artificially large \change. {In this model, no flares were seen at the end of the 500 year model.}
\change\ $>1$ indicates a net increase and \change\ $<1$ indicates a decrease in abundance as a result of the cumulative impact of flares. \change\ $\approx1$ suggests that species are unaffected by 500 years of stochastic flares.

\begin{figure*}
    \centering
    \includegraphics[scale=0.7]{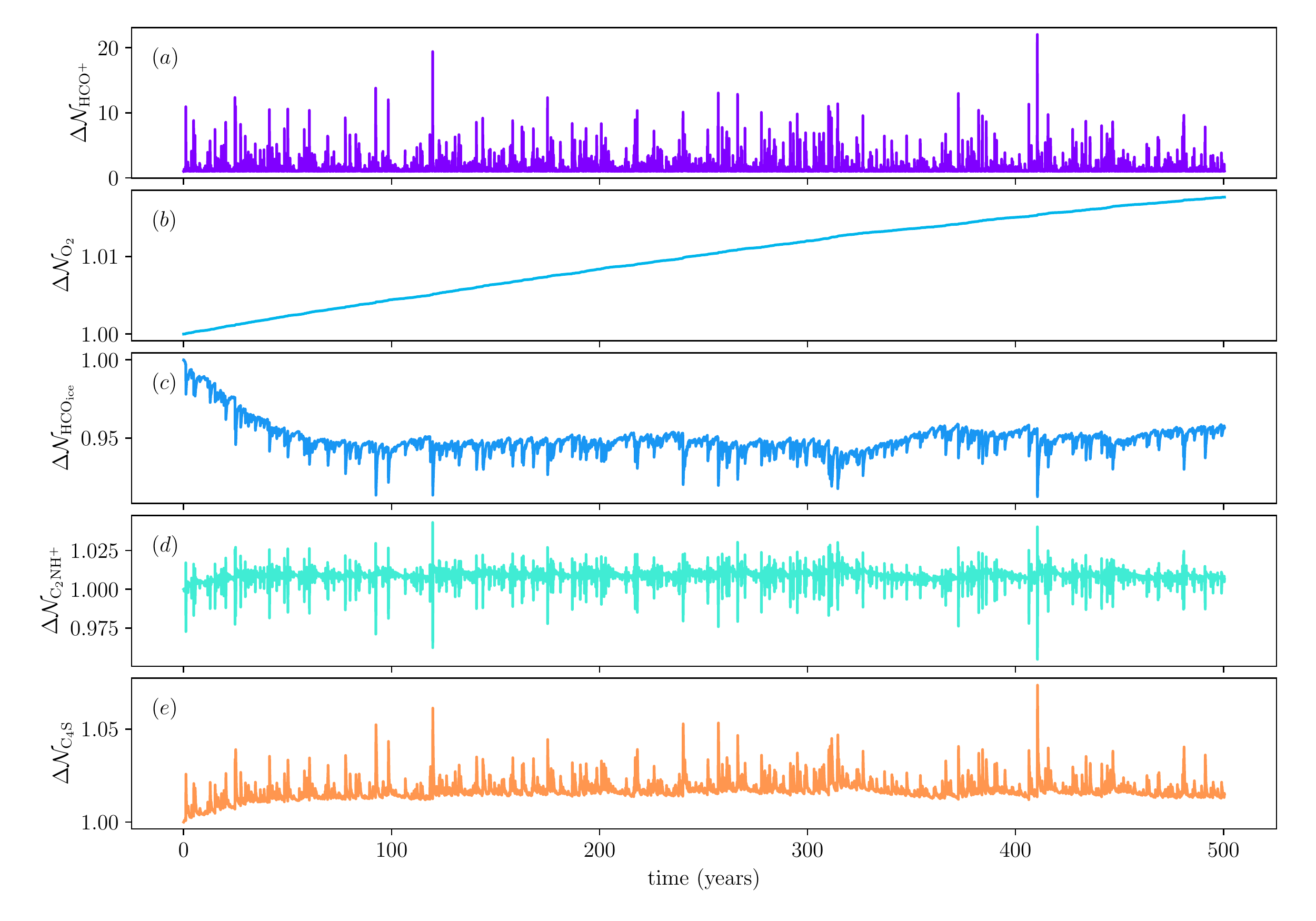}
    \caption{
    Examples of the types of changes seen in disk integrated number, $\Delta$\fancyN, over the model duration.
    (\textit{a}) Certain species, like HCO$^+$, vary in response to individual flares. 
    (\textit{b}) Species, such as O$_2$, are not impacted by individual flares, but exhibit a slow and gradual change in abundance over the course of the model. 
    (\textit{c{,e}}) Species, such as HCO$_{\rm ice}$ {and C$_4$S}, are responsive to individual flares and have a shifted baseline abundance
    (\textit{d}) Species, such as 
    C$_2$NH$^+$ exhibit both enhancement and destruction as a result of flares
    }
    \label{fig:variability_example}
\end{figure*}

{To test the validity of using a fixed time at 0.5 Myr to compute $\sigma$ and \change, we ran an additional 500 year model with a constant X-ray ionization rate (no flares). For the sake of computational efficiency, we ran the evolution after 0.5 Myr in five 100-year time steps rather than the 4-hour time steps in the fiducial model. 
We find the relative change in a given species's abundance in the non-flare model over 500 years is very small, with \fancyN\ changing by less than 1\%. 
This confirms that the chemical model had indeed reached steady state prior to flare initiation and that changes in chemical abundance are caused by variations in the X-ray ionization rate.}

\section{Results}\label{sec_results}

{Given the large number of locations, species, and timesteps considered in the simulations, we have aimed to focus the results by grouping them into types of responses as well as into chemical families.}
{Furthermore, we define here several key terms used throughout the Results and Discussion sections of this work. The disk mid-plane corresponds to temperatures $<20$K, where CO freeze out occurs. The {molecular layer corresponds} to regions where CO can be in the gas-phase ($>20$K). The photo-dissociation or photon-dominated region is defined as regions where CO is dissociated to C and C$^+$.
For more on the physical and chemical structure of disks see the review of \citet{oberg2021}, and references therein.
}

\subsection{General Findings}

Chemical responses to X-ray flaring events fall into three categories, as demonstrated in Figure \ref{fig:variability_example}. {While the following percentages are highly model dependent, they give the reader a baseline to compare which species are more changed compared to other species.}
First are flare sensitive species, which are defined as species with an abundance that varies as a direct result of an individual flare. Flare sensitivity is measured by standard deviation ($\sigma >0.05$),
where the most sensitive species have large $\sigma$ values. 
Second are species with altered `steady-states,' meaning that their average abundance over time has consistently changed over the duration of the model. 
An altered steady state is measured by the relative change in abundance that is at least \change\ $> 5\%$. While a 5\% change may sound like a small percentage, it is possible that over much longer - astrophysically relevant - time scales these species could continue to trend either upwards or downwards. 
Third are species that are non-responsive to flares. These species have $\sigma \le 0.05$ and \change\ $\le 5\%$.

The majority of species in the model are unaffected by flares. $27\%$ of species are considered significantly variable ($\sigma > 0.05$), and $8\%$ have an altered steady state (\change\ $ > 5\%$). Even though a relatively small percent of the total species in the network appear to be impacted by flares, a large fraction of gas-phase species, especially gas-phase cations, are susceptible to flares. Average standard deviations and percent changes of each chemical family are shown in Table \ref{tab:chem_families}. We note that species bearing phosphorus, chlorine, or metals (e.g., Mg, Fe) are excluded from this analysis due to their low gas phase abundances, under the expectation that they primarily take refractory form. 

{The chemical species used in this model can be categorized into chemical families, where families are defined as a group of gas-phase species containing oxygen, carbon, nitrogen, sulphur, or silicon. Additional families include gas-phase anions, cations, and neutrals in addition to ices frozen out on dust grains.
A single species can belong to multiple families. 
Table \ref{tab:chem_families} summarizes the standard deviation and relative change of each of these families.}

\subsection{Cations: The Most Responsive Species}\label{sec:cations}

\begin{figure}
    \centering
    \includegraphics[scale=0.22]{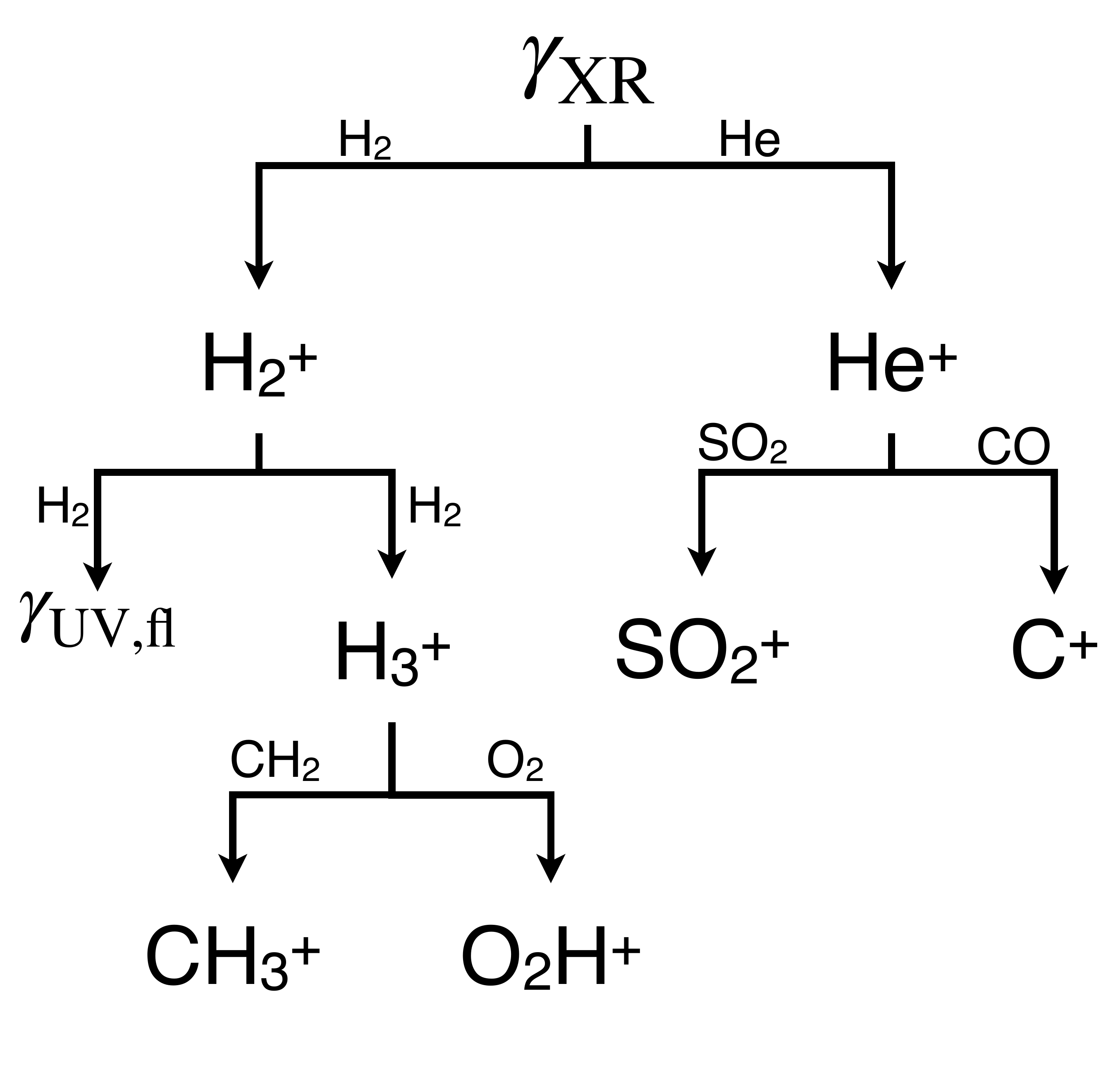}
    \caption{Initial products of X-ray ionization in the disk.
    $\gamma_{\rm XR}$ indicates X-ray photons and $\gamma_{\rm UV,fl}$ are fluoresced UV photons. When a flare occurs, the X-ray flux increases, then increasing the ionization rates of H$_2$ and He and causing a temporary enhancement in gas-phase cations.
    This reaction network is the main driver in all other flare driven chemistry seen in the model.}
    \label{fig:supers_reactions}
\end{figure}

Cations are the most flare-responsive group of species in the model. 
Cations are more likely to experience both an immediate response to flares and are more likely to be impacted on longer time scales ($\gtrsim$ months) as a result of a single, strong flaring event. 
Gas-phase cations make up $39\%$ of the chemical network.
$23\%$ of the modeled cations 
have $\sigma > 0.05$, and 
$19\%$ have \change\ $> 5\%$.

Among the flare sensitive cations, H$_3^+$, H$_2^+$, SO$_2^+$, and O$_2$H$^+$ are significantly more variable than any other species, including ionized and neutral gas-phase species and ices, in the network.
These four species are seen to increase by up to $500\%$ in \fancyN\
in response to the strongest flares ($\Delta L_{\rm XR} \geq 600$). 
The flare enhanced X-ray ionization rate drives a direct enhancement in 
H$_2^+$. H$_2^+$ drives further reactions, leading to the enhancement in H$_3^+$ and O$_2$H$^+$ by 
\begin{equation}
    \rm H_{2} + \gamma_{XR} \rightarrow  H_{2}^{+} + e^{-} \label{h2ion}
\end{equation}
\begin{equation}
    \rm H_2^+ + H_2 \rightarrow H_3^+ + H \label{h2p_h2_h3p_h}
\end{equation}
\begin{equation}
    \rm H_3^+ + O_2 \rightarrow O_2H^+ + H_2. \label{h3p_o2_o2hp_h2}
\end{equation}
Enhancement of SO$_2^+$ is a result of X-ray ionization of helium, rather than H$_2$: 
\begin{equation}
    \rm He^+ + SO_2 \rightarrow SO_2^+ + He.
\end{equation}

An example reaction network of these four species and their immediate products
is shown in Figure \ref{fig:supers_reactions}.
In general, most flare responses can be traced back to the ionization of H$_2$ and He or UV photons ($\gamma_{\rm fl,UV}$) produced by collisional de-excitation of H$_2^+$. {After ionized H$_2$ and He, the dominant chemical drivers are CH$_3^+$ produced through
\begin{equation}
    \rm CH_2 + H_3^+ \rightarrow CH_3^+ + H_2, 
    \label{ch2_h3p_ch3p_h2}
\end{equation}
and C$^+$ from CO:
\begin{equation}
    \rm He^+ + CO \rightarrow C^+ + O + He.
\end{equation}}

\begin{figure}
    \centering
    \includegraphics[scale=0.58]{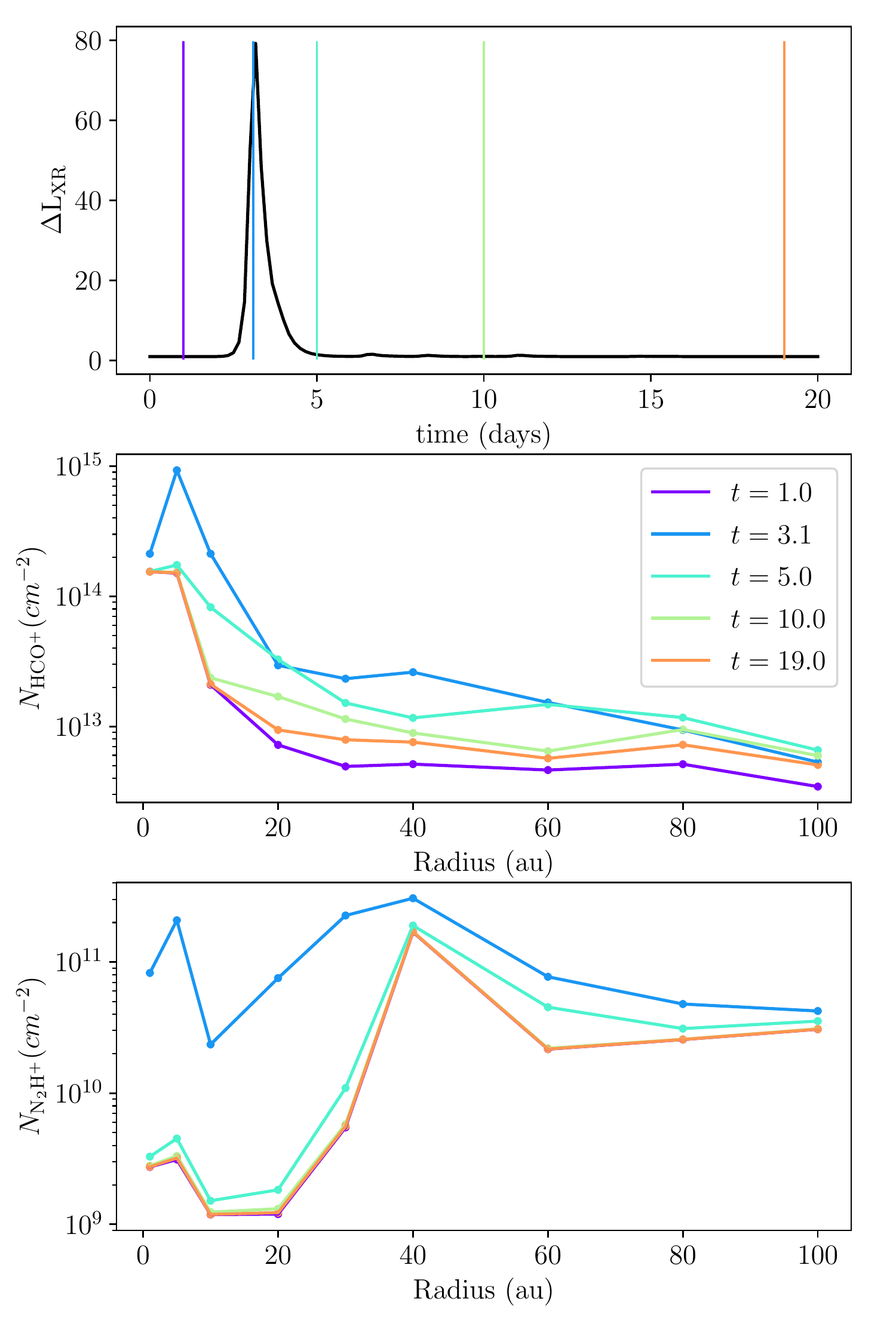}
    \caption{{\textit{Top:} X-ray flare with a strength of $\Delta L_{\rm XR,peak}=80$ (Equation \ref{eq_lxr}). The colored lines correspond to the time steps HCO$^+$ and N$_2$H$^+$ column density if plotted below. 
    \textit{Middle and Bottom:} Column density (\textit{N}) of HCO$^+$ and N$_2$H$^+$ before the flare {($t=1.0$ day)}, at the flare peak {($t=3.1$ day)}, and after the flare {($t=5.0,10.0,19.0$ days)}. Note that both species are enhanced more at shorter radii, and HCO$^+$ is enhanced for a longer time ($\sim 20$ days) than N$_2$H$^+$ ($\sim 5$ days). {N$_2$H$^+$ has the same column density at $t=1.0$, $10.0$, and $19.0$ days.}
    }}
    \label{fig:colden}
\end{figure}

{Of the family of cations impacted by flares, HCO$^+$ and N$_2$H$^+$ are of of primary interest, as they are commonly observed species in disks. Both of these species are temporarily enhanced by flares. Enhancement is a product of protonation from H$_3^+$, where HCO$^+$ is formed by
\begin{equation}
    \rm H_3^+ + CO \rightarrow HCO^+ + H_2, \label{h3p_co_hcop_h2}
\end{equation}
and N$_2$H$^+$ is formed by 
\begin{equation}
    \rm H_3^+ + N_2 \rightarrow N_2H^+ + H_2
\end{equation}
Dissociative recombination with electrons is the primary destruction mechanism for both species.
N$_2$H$^+$ and HCO$^+$ experience a nearly instantaneous enhancement, but the HCO$^+$ enhancement is sustained for longer periods of time. Figure \ref{fig:colden} shows how the radial column density of HCO$^+$ and N$_2$H$^+$ vary as a function of time during a strong flare. HCO$^+$ stays elevated at all radii for longer ($\sim1$ week) than N$_2$H$^+$, which {returns to its pre-flare abundance after a few days}. The reason for this difference is that the N$_2$H$^+$ traces gas closer to the midplane where the densities are generally higher, which speeds up the dissociative recombination.}

\subsection{Neutral Species}\label{sec:neutrals}

Neutral species (both gas and ice-phase) make up $59\%$ of the chemical network, and the vast majority are unaffected by flaring events. Only $0.6\%$ 
of the neutral species have $\sigma > 0.05$, and only $0.9\%$ 
have \change\ $> 5\%$. 
The most variable neutral species are H, HCO, O$_2$H, C$_3$O, C$_2$H$_4$, and C$_3$H$_4$.
In this section, we highlight the chemical and physical processes that drive variability in these species. 

Variability in neutral species is driven by an enhancement in H$_3^+$.
For example, {H$_3^+$ enhances HCO$^+$ (see Reaction \ref{h3p_co_hcop_h2}), which then }
{accepts an electron from a neutral donor (represented by M),} 
\begin{equation}
    \rm HCO^+ + M \rightarrow HCO + M^+, \label{hcop_e_hco}
\end{equation}
to enhance HCO abundance. 
HCO variability occurs primarily beyond $R =5$\,au and within the vertical heights $Z/R = 0.2$ and $0.3$. 
HCO is relatively constant in the disk surface, where UV photolysis and destruction reactions with cations are significantly faster than the neutralization of HCO$^+$. 
Between $Z/R = 0.2$ and $0.3$ HCO enhancement closely follows HCO$^+$ enhancement.
Below $Z/R=0.2$ HCO is not significantly abundant due to the reduced ionization rates. 

Additionally, H$_3^+$ leads to an enhancement in CH$_3^+$ by {Reaction \ref{ch2_h3p_ch3p_h2}}.
CH$_3^+$ then can either react with small oxygen bearing species, such as OH, 
\begin{equation}
    \rm CH_3^+ + OH \rightarrow H_2CO^+ + H_2 \label{ch3p_oh_h2cop_h3},
\end{equation}
or be converted to CH$_4$ through a radiative association step:
\begin{equation}
    \rm CH_3^+ + H_2 \rightarrow CH_5^+, \label{ch3p_h2_ch5p}
\end{equation}
followed by a dissociative recombination step:
\begin{equation}
    \rm CH_5^+ + CO \rightarrow HCO^+ + CH_4. \label{ch5_co_hcop_ch4}
\end{equation}
These reactions and products branch off to drive enhancement in species such as O$_2$H, C$_3$O, C$_2$H$_4$, and C$_3$H$_4$. 
CH$_3^+$ enhancement also plays an important role in carbon, oxygen, and silicon variability
as discussed in later sections. 

We note that H$_3^+$ indirectly leads to a 0.5\% increase in the amount of free hydrogen atoms. However, given the large uncertainties on H$_2$ reformation on grains (including H-binding energies), we do not expect this to significantly change the chemistry. 

\subsection{Chemical Families}

\begin{table}[]
    \begin{center}
        \caption{Average standard deviation ($\bar{\sigma}$) and average change ($|1-\bar{\mathcal{C}}|$) for the defined chemical families. Note: H$_3^+$, H$_2^+$, SO$_2^+$, and O$_2$H$^+$ were excluded from the O and S families in these values, as they {are significantly more variable than any other species in the network and skewed $\sigma$ and $\mathcal{C}$ values. Additionally, phosphorus, chlorine, or metals (e.g., Mg, Fe) are excluded, as they have abundances below the computational limit of the model.}.\label{tab:chem_families}}
    \begin{tabular}{c|c|c|c|c}
        Family  & $\bar{\sigma}_{\rm all}$ & $\bar{\sigma}_{\rm cations}$ &
        $|1-\bar{\mathcal{C}}_{\rm all}|$ &
        $|1-\bar{\mathcal{C}}_{\rm cations}|$ \\
        \hline
       Cations & ---   & 0.13  & ---   & 0.03  \\ 
        Anions  & 0.02  & ---   & 0.01  & ---   \\
        Neutrals& 0.00  & ---   & 0.00  & ---  \\
        \hline
         O      & 0.06  & 0.10  & 0.03  & 0.05  \\ 
         S      & 0.05  & 0.08  & 0.04  & 0.06  \\
         C      & 0.02  & 0.03  & 0.01  & 0.02  \\
         N      & 0.02  & 0.04  & 0.01  & 0.02  \\
         Si     & 0.01  & 0.2   & 0.01  & 0.02  \\
        Ices    & 0.00  & ---  & 0.00  & ---  \\
    \end{tabular}
    \end{center}
\end{table}

\subsubsection{Oxygen Bearing Species}\label{sec:ospecies}

\begin{figure}
    \centering
    \includegraphics[scale=0.22]{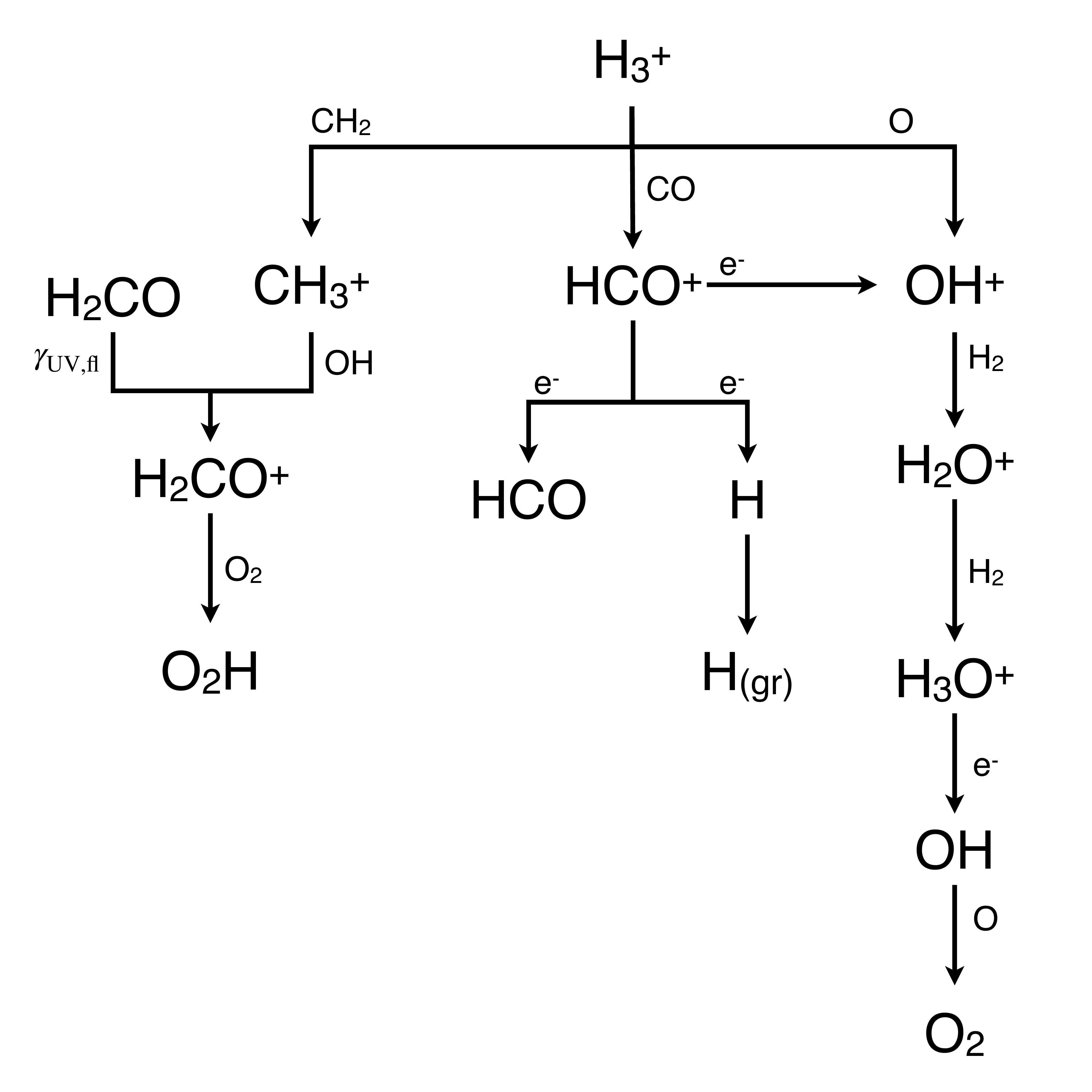}
    \caption{Example flare-driven reaction network of oxygen bearing species. O-bearing species are the most flare susceptible chemical family in the network. The strongest responses occur {in regions} of the disk near the star and disk surface.
    X-ray flares enhance H$_3^+$ in the disk, which then protonates small gas-phase neutral species, such as CO, O, and CH$_2$. Some O-bearing species, such as O$_2$H and HCO$^+$ are variable on short time scales (days-weeks), whereas others, such as O and O$_2$, are gradually impacted over the 500 year model.
    }
    \label{fig:oxygen_reactions}
\end{figure}

Oxygen bearing species are the most variable chemical family, where 
$29\%$ of gas-phase O-bearing species have $\sigma > 0.05$, and $22\%$ have \change\ $>5\%$.
Variability among these species tends to occur in warmer disk regions, i.e., dominantly near the disk surface ($Z/R>0.3$) and close to the star ($R<40$\,au).
For the most part, changes in O bearing species can be traced to two pathways originating from enhancement of H$_3^+$, as summarized by Figure \ref{fig:oxygen_reactions}.

One chemical pathway is driven by enhancement of CH$_3^+$ (Reaction \ref{ch2_h3p_ch3p_h2}). CH$_3^+$ reacts with small species, such as OH, to form larger C and O bearing species, such as H$_2$CO$^+$. This reaction branch drives short-term variability in larger O bearing species, including O bearing carbon chains. 
Additionally, oxygen chemistry is driven by H$_3^+$, which directly protonates small O bearing species, such as CO and O. 
This reaction branch is interesting, as protonation drives both short-term variability and long-term changes in abundance. 
Temporary enhancement of HCO$^+$, as discussed in Section \ref{sec:neutrals}, is an example of H$_3^+$ driven short-term variability. 

Flares are also seen to slowly convert {atomic} O to O$_2$ over the 500 year model.
The dominate process occurs via a series of gas-phase protonation reactions to form H$_3$O$^+$: 
\begin{equation}
    \rm H_3^+ + O \rightarrow H_2 + OH^+ \label{h3p_o_h2_ohp}
\end{equation}
\begin{equation}
    \rm OH^+ + H_2 \rightarrow H_2O^+ + H
\end{equation}
\begin{equation}
    \rm H_2O^+ + H_2 \rightarrow H_3O^+ + H. \label{h2op+h2_h3op_h}
\end{equation}
H$_3$O$^+$ then undergoes dissociative recombination to form OH,
\begin{equation}
    {\rm H_3O^+ +} e^- \rightarrow {\rm OH + H_2}.
\end{equation}
{Additionally, H$_3$O$^+$ can form H$_2$O, as discussed in \citet{waggoner2019}}.
OH can then undergo a substitution reaction with O to form O$_2$
\begin{equation}
    \rm OH + O \rightarrow O_2 + H.
\end{equation}

\fancyN$_{\rm O, {atomic}}$ decreases by $0.2\%$, and \fancyN$_{\rm O_2}$ increases $1.2\%$ over the 500 year model. 
While these changes may seem insignificant, {atomic} O and O$_2$ are both abundant species in the disk 
(O and O$_2$ abundances are $1.96 \times 10^{-6}$ and $1.31 \times 10^{-7}$ with respect to H, respectively). {Since \change\ is a relative change in disk integrated abundance abundant species, such as O$_2$ and especially O, are less likely to exhibit a large change in \fancyN.} 
{Therefore, these seemingly small fractional changes are indicative of a non-insignificant conversion of O to O$_2$, as further discussed in Section \ref{sec:perma} and shown in Figure \ref{fig:variability_example}b.
}

\subsubsection{Carbon Bearing Species}

\begin{figure}
    \centering
    \includegraphics[scale=0.22]{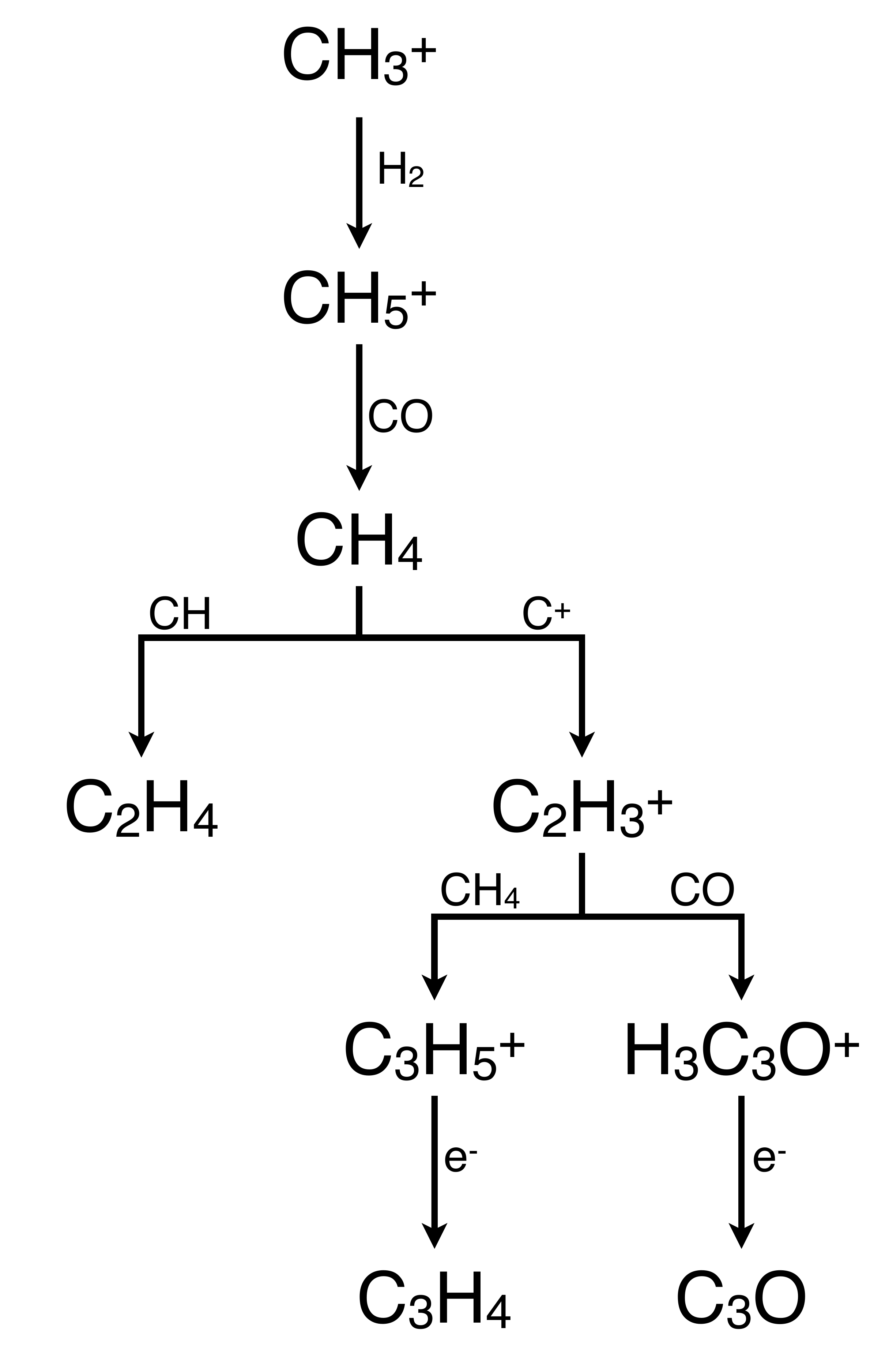}
    \caption{Example flare-driven reaction network of carbon bearing species. C-bearing species are the second most flare susceptible chemical family in the model, where the strongest responses occur in colder regions of disk. 
    Variability in C-bearing species, especially carbon chains, dominantly come from flare enhanced production of CH$_4$ and C$^+$. 
    }
    \label{fig:carbon_reactions}
\end{figure}

Gas-phase carbon bearing species are the third most variable species, behind sulphur bearing species. 
While S-bearing species have higher $\sigma$ and \change\ values than C-bearing species on average, it should be noted that S-bearing species are generally less abundant and are therefore more susceptible to small changes in abundance than C-bearing species. 
Gas-phase C-bearing species make up $52\%$ of the chemical network, where
$9\%$ of gas-phase C-bearing species have $\sigma > 0.05$, and $8\%$ have \change\ $>5\%$.

As discussed in Section \ref{sec:neutrals}, carbon chain chemistry can be traced to enhancement of C$^+$ and CH$_3^+$. While enhanced, C$^+$ and CH$_3^+$ drive carbon protonation reactions and neutral-neutral and neutral-ion combination reactions between different carbon bearing species, as demonstrated in the enhancement of C$_3$H$_4$ and C$_2$H$_4$ (Figure \ref{fig:carbon_reactions}). 
Typically, C$^+$ tends to drive variability closer to star and along the disk surface. CH$_3^+$ tends to drive variability in carbon chemistry along the warm molecular layer. 
As a whole, the cumulative impact of flares tends to produce larger neutral and ionized carbon chains. 

\subsubsection{Sulphur Bearing Species}

\begin{figure}
    \centering
    \includegraphics[scale=0.22]{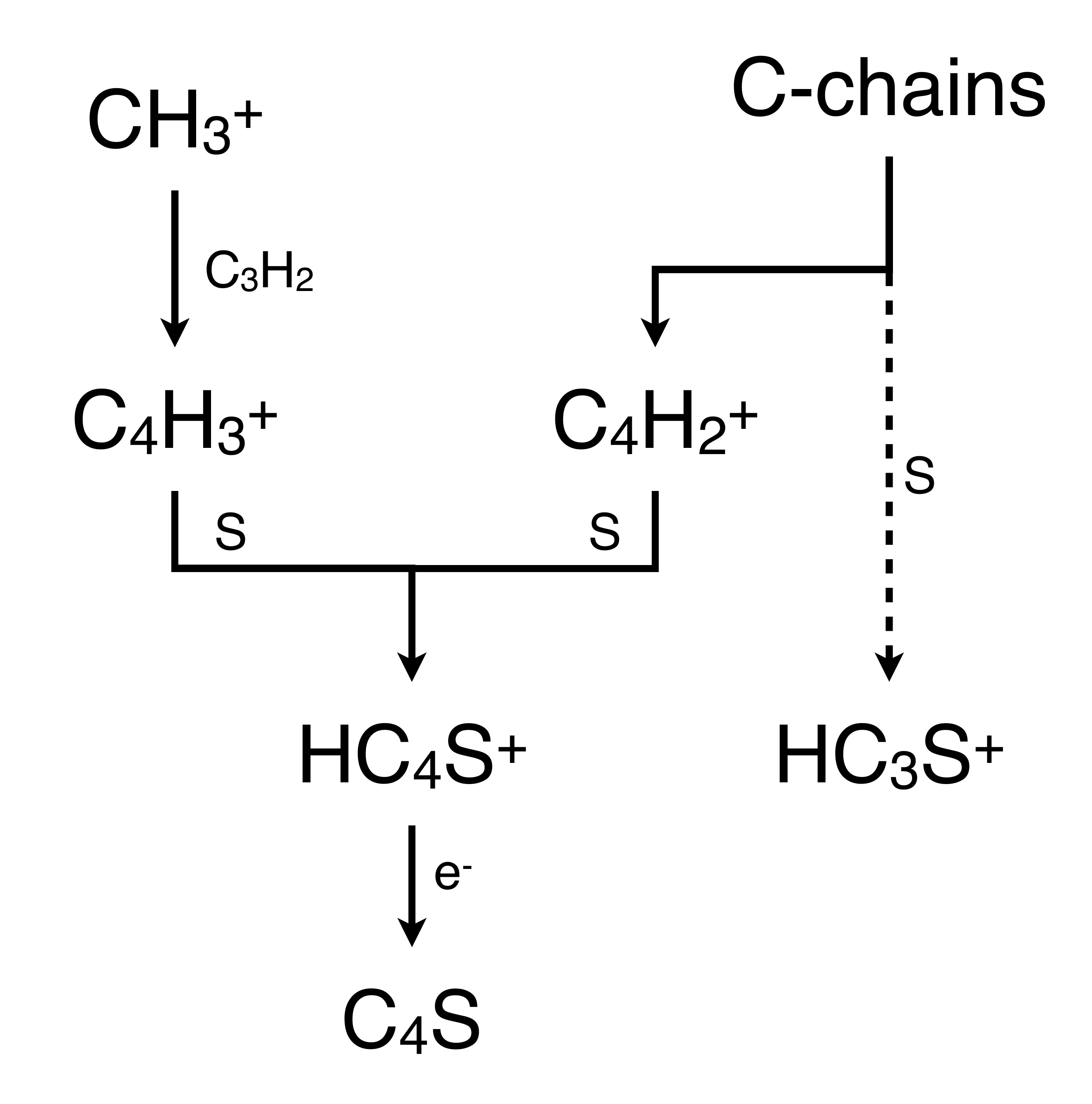}
    \caption{Example flare driven reaction network for sulphur bearing species. Variability in S-bearing species is tied to the carbon chain chemistry (Figure \ref{fig:carbon_reactions}), where carbon chain cations can react with small S-bearing species (such as S) to form larger organosulphur species. 
    }
    \label{fig:sulphur_reactions}
\end{figure}

Gas-phase sulphur bearing species make up $8\%$ of the chemical network, where $18\%$ have $\sigma > 0.05$ and $16\%$ have \change\ $> 5\%$.
Variability in S-bearing species is directly tied to flare driven carbon chain chemistry, as demonstrated in Figure \ref{fig:sulphur_reactions}.
C$^+$ and CH$_3^+$ lead to variability in carbon chains, which react with atomic sulphur and other small S-bearing species, such as H$_2$S$_2$, H$_2$S, OCS, and S$_2^+$,
to form S-bearing organics. 
These smaller S-bearing species are slowly and gradually converted to organosulphur species, most notably C$_4$S, over the 500 year model at the few percent level (e.g., \change$_{\rm S} = 0.99$ and \change$_{\rm C_4S} = 1.02$).

\subsubsection{Nitrogen Bearing Species}

\begin{figure}
    \centering
    \includegraphics[scale=0.22]{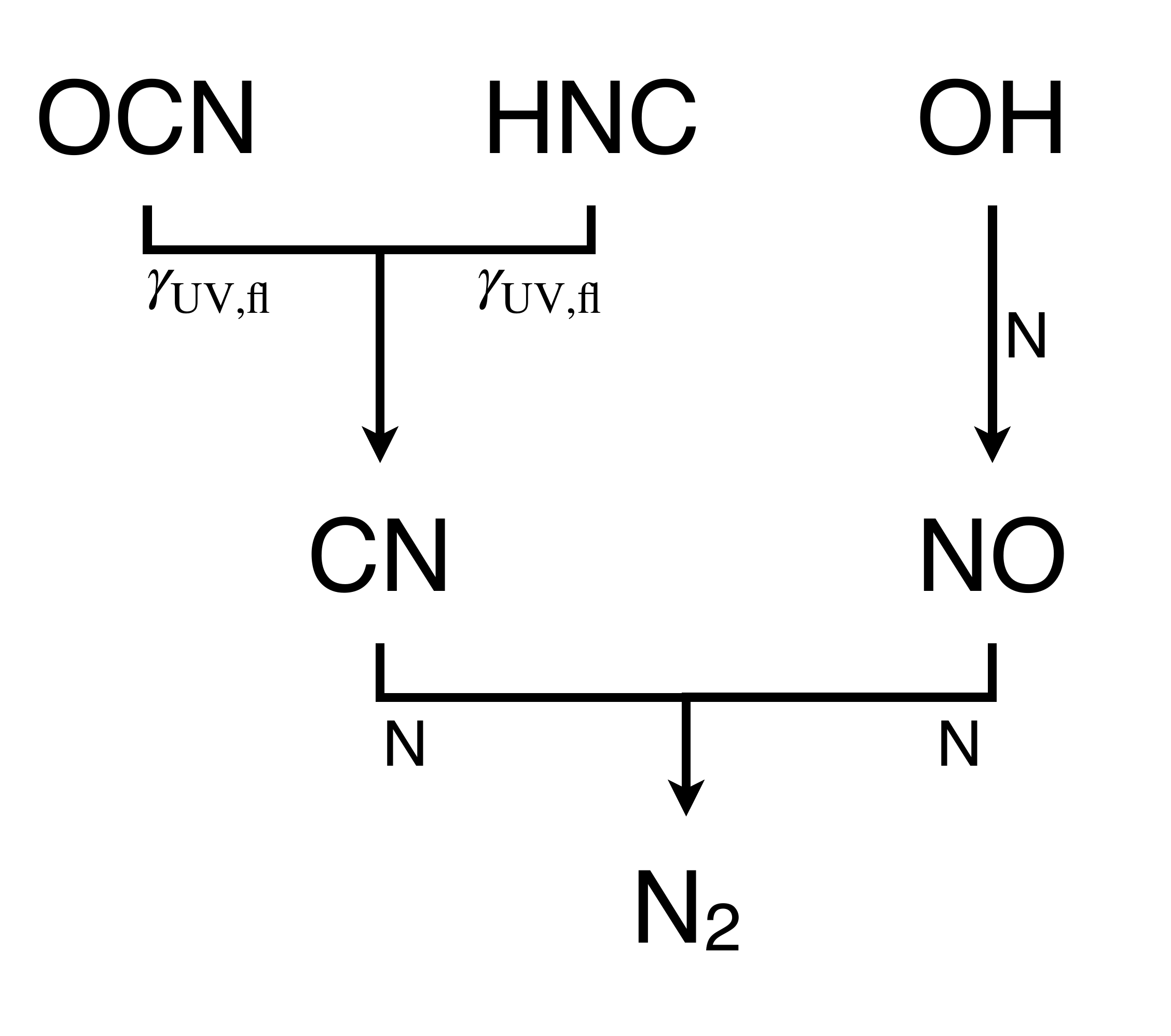}
    \caption{Example flare driven reaction network for nitrogen bearing species. N-bearing species are among the least variable chemical family, but the model suggests that flares steadily convert {atomic nitrogen to
    N$_2$. }
    }
    \label{fig:nitrogen_reactions}
\end{figure}

\begin{figure*}
    \centering
    \includegraphics[scale=0.65]{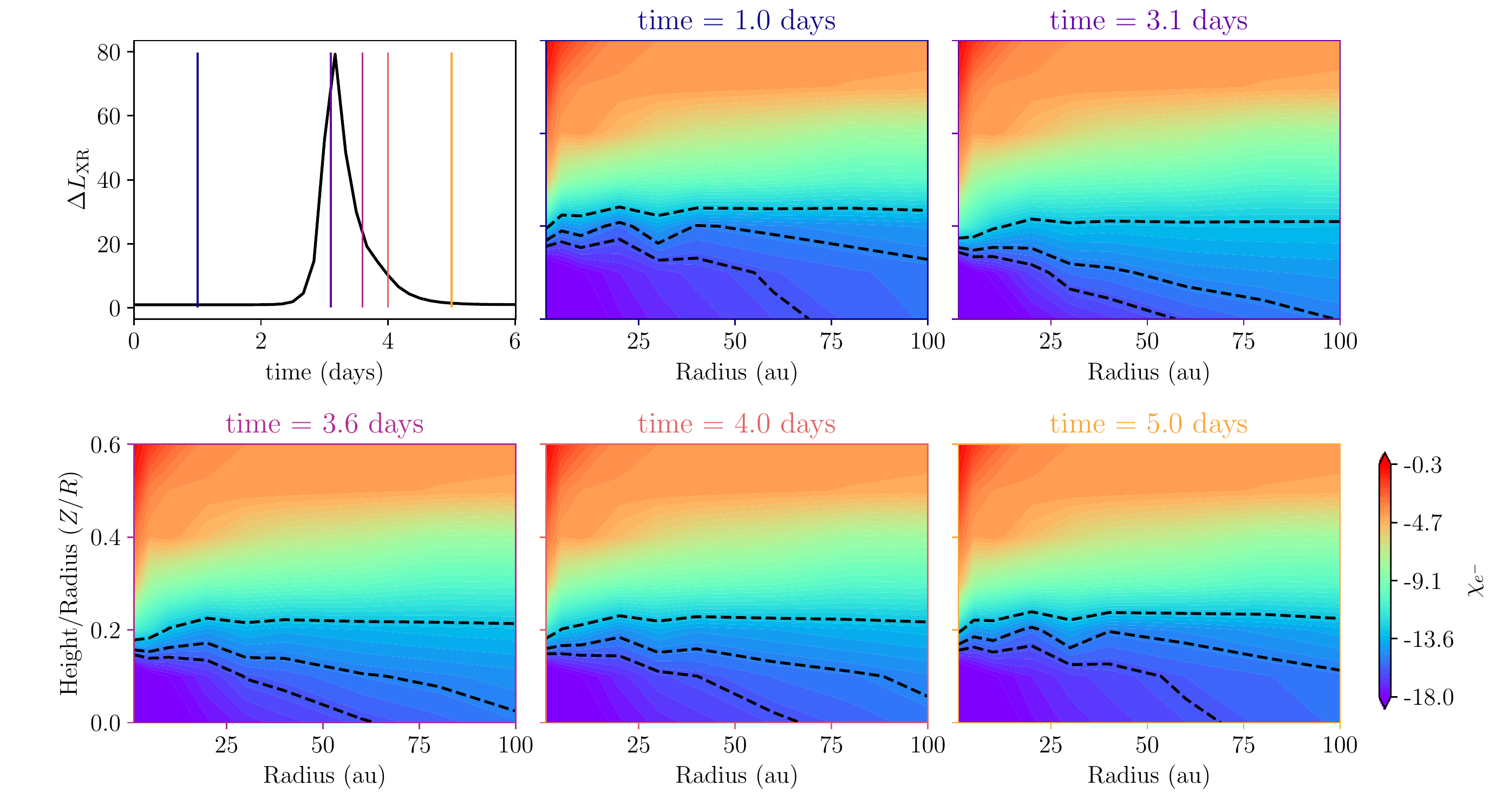}
    \caption{Fractional electron abundance ($\chi_{e^-}$) with respect to total H in response to an X-ray flare. An X-ray flare with a peak $\Delta L_{\rm XR}=80$ (top left) temporarily enhances $\chi_{e^-}$ {below $Z/R=0.2$} during the flare ($t=3.1$ and $3.6$\,days). After the flare has ended ($t=5.0$\,days),  $\chi_{e^-}$ returns to its pre-flare abundance ($t=1.0$\,day). The colors of the axes denote the time steps labeled by the vertical lines in the top left panel.  
    Within each sub-panel of electron abundance, the black contour line indicates $\chi_{e^-}=10^{-16}$, $10^{-15}$, and $10^{-13}$, for reference. 
    }
    \label{fig:ion}
\end{figure*}

\begin{figure}
    \centering
    \includegraphics[scale=0.6]{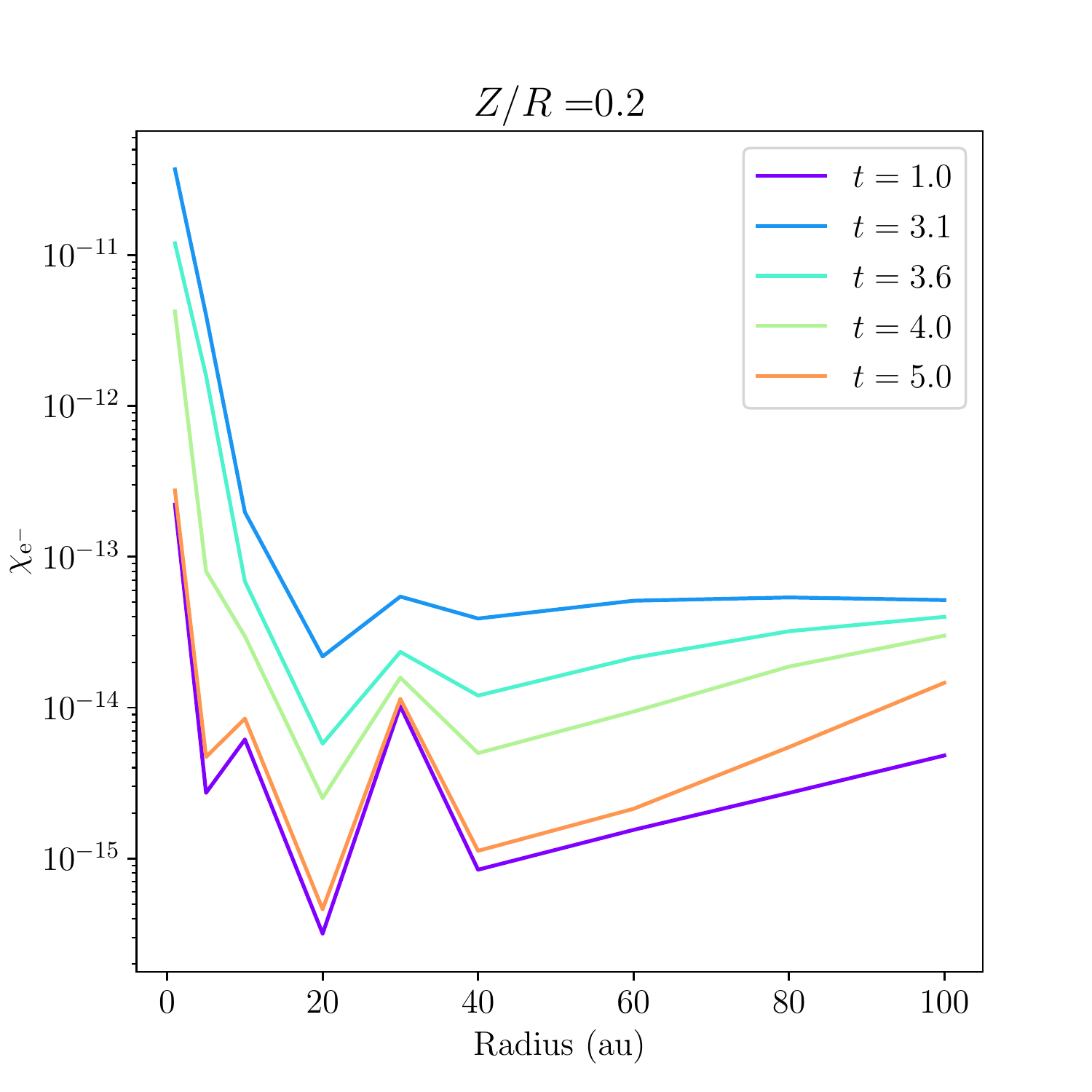}
    \caption{Fractional electron abundance with respect to H ($\chi_{e^-}$) along the vertical height $Z/R=0.2$ for the flare shown in Figure \ref{fig:ion}. Times are plotted in days at the same steps as shown in Figure \ref{fig:ion}. 
    The flare peak occurs at $t=3.1$\,days.
    Electron abundance increases by more than an order of magnitude as a result of the flare. 
    $\chi_{e^-}$ returns to its pre-flare abundance once the flare had ended. }
    \label{fig:eden_zr0.2}
\end{figure}

The majority of species containing nitrogen are impacted on smaller scales than the other chemical families. 
Most of the nitrogen in the model is in the form of N$_2$, with fractional amounts in species like N, CN, HCN, and NH$_3$. 
{While the average $\sigma$ and \change\ values for N-bearing species (see Table \ref{tab:chem_families}) suggest that N-bearing species are variable similarly to C-bearing species, we note that N-bearing species {typically have lower number densities} (N$_{\rm tot} = 7 \times 10^{-5}$ w.r.t. H{$_{\rm tot}$}, C$_{\rm tot} = 1 \times 10^{-4}$ w.r.t. H{$_{\rm tot}$}). Therefore, the relative variability naturally reports larger than for more abundant species.}
Gas-phase N reacts with CN and NO (see Figure \ref{fig:nitrogen_reactions}), resulting in the formation of N$_2$.
Both CN and NO are increased in abundance in response to individual flaring events, thus temporarily increasing N$_2$ production during flares. Nitrogen is then locked up in N$_2$, and over the course of the 500 year model CN, NO, and N have a decreasing average abundance over time due to cumulative flares. 
While the percent change in disk integrated abundance of nitrogen bearing species
is small, we note that these are also relatively abundant species. 

\subsection{Electron Density}\label{sec:eden}

The disk-integrated number of electrons
(\fancyN$_{e^-} = 7.8 \times 10^{47}$)
is relatively constant throughout the 500 year model ($\sigma < 0.01$, \change $< 1\%$) \footnote{Note that the disk overall remains ``charge neutral'' as required by the model.}. However, the fractional electron abundance with respect to total H ($\chi_{e^-}$) in the warm molecular layer appears to be sensitive to discrete flaring events.
{We find that photoionization from both X-rays and fluoresced UV photons is the main mechanism behind electron enhancement. Dissociative recombination with positively charged molecular ions was found to be the primary mechanism that brings electrons back to their pre-flare abundance.}
Figure \ref{fig:ion} shows the time variable $\chi_{e^-}$ in the disk before a flare, near the end of a flare, and after the flare.
While $\chi_{e^-}$ in the upper disk surface and mid-plane are constant ($\Delta \chi_{e^-} \approx 1$) during a flare, electron abundance at intermediate heights {($Z/R<0.2$)} can be temporarily enhanced by several orders of magnitude, as shown in Figure \ref{fig:eden_zr0.2}. The surface is less susceptible since the electron abundance is dominated by UV photo-ionization. The midplane is also less susceptible due to the impact of flares dropping closer to the dense, X-ray attenuated midplane.  
{Additionally, enhancement lasts for longer time scales at large radii, since dissociative recombination timescale scales with ion density. Therefore, more ionized regions, such as those close to the central star, typically have faster recombination rates. Recombination scales with respect to HCO$^+$ chemistry are explored more in depth in \citet{cleeves2017}.}

{The} local electron abundance is sensitive to individual flares, but the disk integrated electron abundance remains effectively constant throughout the model. Therefore, there does not seem to be a long-term cumulative impact of flares on disk ionization.

{\subsection{Significance of Specific Flare History}}

\begin{figure*}
    \centering
    \includegraphics[scale=0.75]{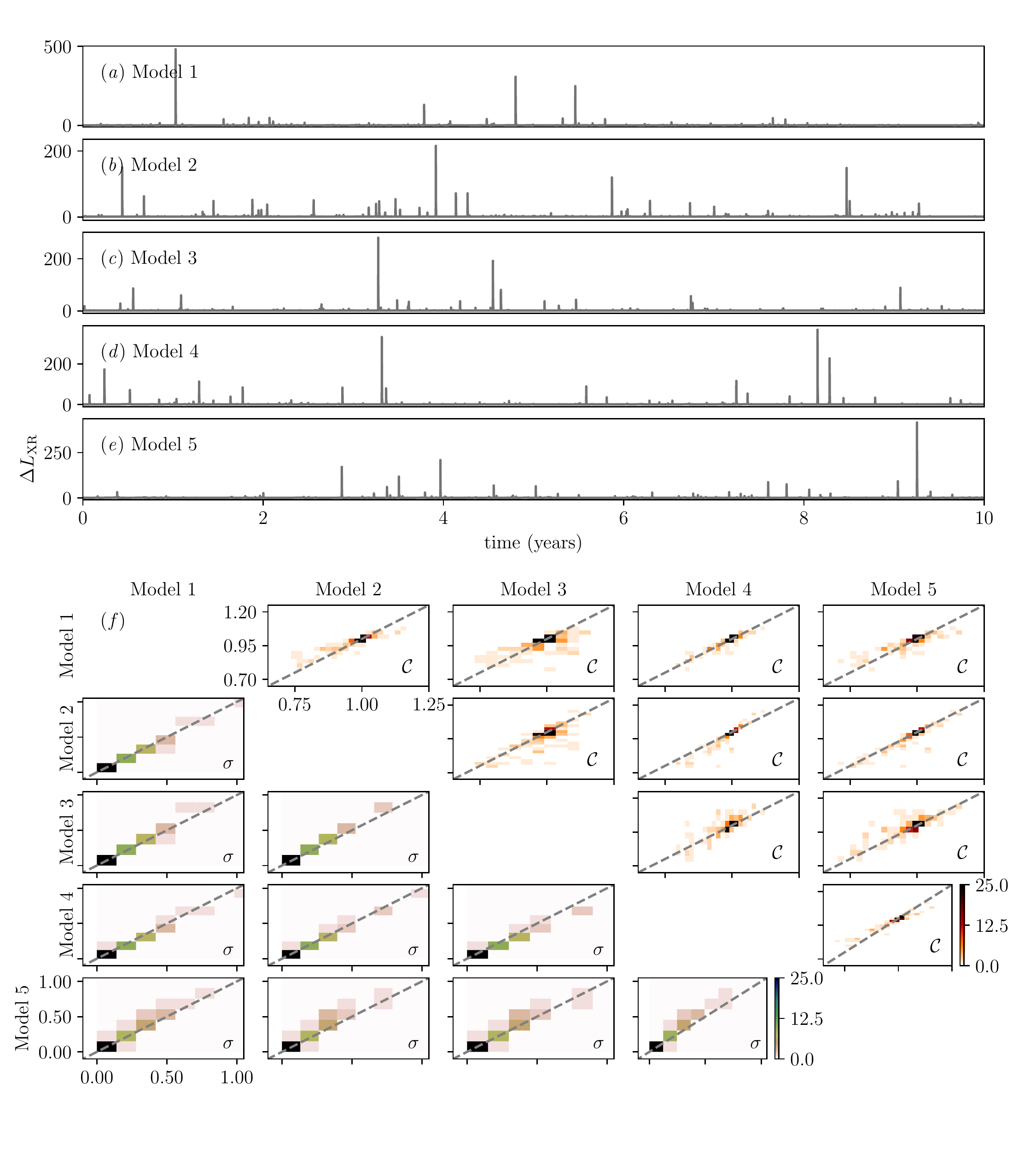}
    \caption{(\textit{a-e}) The first ten years of the X-ray light curve for all five 100 year light-curve models.  
     (\textit{f}) A comparison plot of standard deviation ($\sigma$, bottom left plots) and relative change in abundance (\change, top right plots) for each combination of model seeds. All abundant molecules are plotted as a ``heat map'' where the density of species within a point are indicated by the color bar. The different seeds give similar chemical behavior, falling along the indicated one-to-one correspondence line (dashed), 
     indicating that the history of individual flares is not the primary driver of the variations found in the models. 
     }
    \label{fig_seedvals}
\end{figure*}

To test our results' dependence on the specific history of stochastic flares, we ran five additional 100 year models to compare chemical responses to different light curves (Figure \ref{fig_seedvals}a-e). 
Each model was run with identical parameters, save for the randomly generated light curve, where each model was given a different random number seed. 
The five light curves are in agreement with the target energy distribution and frequency, as described in Section \ref{sec:xgen_ttauri}.

The standard deviation and relative change of chemical species for all five models are cross compared in Figure \ref{fig_seedvals}f. 
We find that the standard deviation and the relative change in abundance for each species are consistent across the five models within $\lessapprox 5\%$ of each other.
{We find that across the five models when examined over the full simulation timeframe,} highly variable species are always highly variable, non-responsive species are always non-responsive, and so on. This suggests that {when examined} over longer time scales (e.g., 100 yrs), {the general behavior of} X-ray flare driven chemistry is not dependent on the history of the light-curve, so long as the flare energy distribution and flare frequency of the star are consistent. 

\section{Discussion}\label{sec_disc}

\subsection{Flare Driven Chemistry}

The majority of chemical responses to X-ray flares can be traced back to the ionization of H$_2$ or He, where the immediate byproducts (e.g., $\gamma_{\rm UV,fl}$, H$_3^+$, CH$_3^+$, C$^+$ shown in Figure \ref{fig:supers_reactions}) are responsible for causing most chemical variability in the disk.
In general, X-ray flares are most likely to have a significant impact on small, less abundant (\fancyN\ $< 10^{40}$, or a fractional abundance $< 10^{-16}$ w.r.t. H{$_{\rm tot}$}) cations, but it is possible that any species directly related to H$_2$ and He ionization is susceptible to variable X-ray ionization rates. 
More abundant species (\fancyN\ $> 10^{45}$, or a fractional abundance {$> 10^{-11}$ w.r.t. total hydrogen)} and neutral species typically have a lower global response to flares. However, these species can have strong responses to flares at individual disk locations, as demonstrated by \citet{waggoner2019} in the case of water. 

For the most part, oxygen bearing species, sulphur bearing species, and carbon bearing species are the most flare susceptible species seen in the model (Table \ref{tab:chem_families}). 
Flare driven variability in S-bearing species is directly dependent on variability on the carbon chemical network, so chemical variability in the disk can be broadly categorized as stemming from the oxygen chemical network or the carbon chemical network.

\subsection{{ Application to} the Missing Sulphur Problem}\label{sec:dis_sulphur}

Sulphur bearing species are the second most responsive chemical family in the network. 
This behavior is tied to carbon cation chemistry. Small S-bearing species slowly react with flare enhanced carbon cations to form larger organosulphur compounds. The most notable conversion is from atomic sulphur (\change$_{\rm S} = 0.99$) to C$_4$S (\change$_{\rm C_4S} = 1.02$), as shown in Figure{s \ref{fig:variability_example}e and} \ref{fig:sulphur_reactions}.
{Sulphur abundances are known to be low in disks, and the lower the abundance the more susceptible a species can be to having its abundance altered by flaring events.}
This result is significant, because observations suggest there is a `missing sulphur problem' in the interstellar medium, and disks specifically \citep[e.g.][]{ruffle1999,kama2019,legal2019}. 
Recent literature has suggested that sulphur is locked up in larger and more difficult to detect species in protoplanetary disks and/or frozen out in ices \citep[e.g.][]{laas2019,shingledecker2020}.
The results presented in this work are significant, because they suggest that X-ray flaring events impact the abundance of organosulphur species.

We note that the results in this section are speculative, and should not be taken as definitive. To better constrain the importance of flares on the long-term sulphur chemical network, more comprehensive models will need to be run.
For example, this model is initialized with a reduced sulphur abundance to take into account refractory sulfur forms 
($10^{-9}$ per H{$_{\rm tot}$}) 
compared {to} the Sun and ISM {\citep[$10^{-5}$ per H$_{\rm tot}$,][]{lodders2003}}.

\subsection{Disk Ionization}

Small, gas-phase cations are the most flare susceptible species in the network. Flares enhance the abundance of specific cations both temporarily (i.e., in response to a single flare) and for the duration of the simulation (i.e., reaching a new steady state). However, the disk integrated number of electrons and number of anions is relatively unaffected by flares.
{While the disk integrated number of electrons (\fancyN$_{e^-}=7.8 \times 10^{47}$ at 0.5 Myr) is equal to the disk integrated number of \textit{total} cations in the disk, \fancyN$_{e^-}$
is several orders of magnitude higher than the even most abundant individual cation species in the disk (\fancyN$\sim 10^{43}$ at 0.5 Myr).} Since variability reported in this work 
 is measured by a \textit{relative} change in abundance, {cations tend to have higher reported variability than electrons due to their lower individual abundances. The model, by construction, remains ``charge neutral'' when considering all positively and negatively charged species}.

While the disk integrated number of electrons is relatively constant in time, electron abundance does not respond to flares uniformly at all locations. 
For example, the electron density is relatively constant in the disk surface and mid-plane, but electron abundance can be temporarily enhanced in the warm molecular layer of the disk as a result of a strong flare (i.e., $\Delta L_{\rm XR,peak}>70$).
Figure \ref{fig:ion} demonstrates that such a flare can increase the fractional electron abundance with respect to H by up to several orders of magnitude along the warm molecular layer. For example, Figure \ref{fig:eden_zr0.2} shows that $\chi_{e^-}$ can increase from $10^{-15}$ to $10^{-13}$ during a flare. The lack of change closer to the surface is likely because ionization and electron production there is dominated by UV photolysis of species like carbon, which is approximately two orders of magnitude faster than the X-ray ionization. So, even when X-ray ionization rates are increased by flaring events, UV ionization still dominates at the surface. Similarly, cosmic rays dominate ionization in the disk mid-plane, where gas densities are high enough to block X-ray photons {(see Figure \ref{fig:xrate})}. 

A fractional electron abundance of $\chi_{e^-}=10^{-14}$ has been linked to the minimum ionization required to couple magnetic fields to neutral gas \citep{igea1999}. Once coupled, magnetic fields aid in disk accretion via processes such as magnetorotational instability \citep[MRI;][]{balbus1991}.
While further modeling is required to confirm this, it is possible that disk accretion rates through the molecular layer could be temporarily increased during X-ray flaring events.

{Previous modeling by \citet{Ilgner2006flares} has explored this idea, demonstrating that X-ray flares increase electron column density in magnetically active zones. The focus of that work differs compared to this work in that they analyze the impact of flares on the magnetohydrodynamic evolution of the disk rather than the chemical evolution. Moreover, they focus their modeling between radii of 0.1 au to 10 au. 
They found that an X-ray flare 100 times the characteristic X-ray ionization rate is seen to increase the electron abundance {by a factor of $\sim10$ at a radial distance of 1 au at the vertical ``transition region'' (where the magnetic Reynolds number is 100) for a model assuming low metal abundances. Our models show that the vertical transition occurs around $z\sim0.18$~au, and at this location, a factor of $\sim100$ flare can change the electron abundance by significantly more, over $400\times$ from a baseline of $\chi(e^-)=7\times10^{-15}$. The main reason for this difference is most likely a differing underlying model for the disk and for the X-ray ionization. Their model contains 20\% less mass than ours inside of 10~au, and solves directly for hydrostatic equilibrium, resulting in a vertically ``puffier'' disk {with lower line of sight column densities to the midplane}. 
These two factors make their model more transparent to X-rays, and would result in a higher baseline electron abundance in their model compared to ours. 
We also include Monte Carlo X-ray transport (including scattering) while they consider line of sight absorption of X-rays only \citep{ilgner2006paperI}. They do not explicitly report their baseline electron abundance, but these factors point to their disk being more ionized than ours, resulting in smaller relative changes for similar flare energies.}
This comparison suggests that the specifics of the impact of flares on magnetically active regions in disks will depend on a number of factors, including but not limited to the disk model itself and the treatment of X-rays \cite[see also the detailed discussion of][]{Ilgner2006flares}, but together make it clear that flares have the potential to impact MRI through variable disk ionization fraction, and should be further explored.}

\subsection{``Permanently'' altered species}\label{sec:perma}

An individual flare seems to result in a temporary change in disk-chemistry, but given sufficient time, the system tends to return to the pre-flare chemical state. This suggests that a single flare is unlikely to have a long term impact for  global chemistry. 
However, when the system is exposed to multiple stochastic flares over centuries or longer, the chemical system is unable to fully return to the pre-flare steady state.  
We find that the cumulative impact of many flares drives some species to a ``new pseudo steady state'' in disk chemistry { if the formation or destruction timescales are much longer than the timescales between significant flares. For example, if a species' formation is enhanced as a result of a flare-produced product, but the destruction timescale (via photons or reactions) is greater than the timescales between significant flares, then the system will not have time to relax before the next flare occurrence. The exact timescales will depend sensitively on the molecule of interest and the local physical conditions.}
While the results from our model generally suggest that most {\em long-term} impacted species have a relative change in \fancyN\ of a few percent or less, some of the impacted species do not appear to reach a plateau by the end of the 500 year model. { Therefore it is possible that the full extent of flares in driving chemical change may be much greater.} 

Examples of species exhibiting the latter behavior are O and O$_2$, {where neither O or O$_2$ reach a pseudo-steady-state after flares are initiated.} O is slowly converted to O$_2$ (Section \ref{sec:ospecies}) {at disk positions of $R \geq 20$au and $0.4 \leq Z/R \leq 0.5$}.
The O$_2$ to total gas-phase oxygen ratio (O$_2$/O$_{\rm tot}$) increases by $0.004\%$ (at $t=0$, O$_2$/O$_{\rm tot} = 0.289\%$), while the O to total gas-phase oxygen ratio (O/O$_{\rm tot}$) decreases by $0.01\%$ (at $t=0$, O/O$_{\rm tot} = 2.15\%$). 
This conversion may seem insignificant, but if the conversion continues linearly for 1 Myr in extension to the modeled 500 years, then O$_2$/O$_{\rm tot}$ will increase by $\sim 8\%$, and O/O$_{\rm tot}$ will decrease by $\sim 20\%$. {However, linear extrapolation is an extreme case; in reality the O to O$_2$ conversion rate may slow down or reach a new steady state prior to 1 Myr.}
{Previous work has shown that O$_2$ abundance in comets 67P/Churyumov-Gerasimenko \citep{bieler2015} and 1P/Halley \citep{rubin2015} is significantly higher than laboratory and modeling predictions \cite[e.g.][]{taquet2016,eistrup2019}. While further modeling is required to know the extent at which flares enhance O$_2$, it is possible that flares could contribute to the unusually high O$_2$ abundances seen in comets. 
}

{Species other than O and O$_2$, such as C$_4$S and S (see Section \ref{sec:dis_sulphur}), are seen to exhibit a similar behavior in the model. Table \ref{tbl:highestchanges} in Appendix \ref{apA} contains the top 30 species with the highest \change\ values.}

\section{Conclusions}\label{sec_conc}

We present the first study of the impact of stochastic X-ray flaring events on the long term (500 year) chemistry of a protoplanetary disk. 
The disk chemistry was modeled 
at disk locations ranging from the mid-plane to the disk surface ($0\leq Z/R \leq 0.6$) and from radii spanning $R=1$\,au to $R=100$\,au. We find that X-ray flares have the following impact on disk chemistry:
\begin{enumerate}
    \item X-ray flares can cause a variety of changes to molecular abundances, including short-term (days to weeks), long-term (centuries), or no change on different species in the disk.
    \item {The majority of species strongly impacted by flares are of relatively low abundance. Abundant species and commonly observed species, such as CO, HCN, HNC, C$_2$H, CN, etc., are only impacted up to $\leq 1\%$ by flares.}
    \item Small, gas-phase cations are impacted significantly more than neutral species and ices. Cations are the most likely to be impacted by individual flaring events (i.e., experience short-term variability). This suggests that detections of gas-phase cations (e.g., HCO$^+$, N$_2$H$^+$) will be more reliable if observed multiple times.
    \item Gas-phase {oxygen} bearing species are the most variable chemical family seen in the model, followed by {sulphur} bearing species and {carbon} bearing species. This suggests that X-ray flares { may} play a role in the formation of more complex biologically relevant molecules. However, further modeling { including a more extensive chemical reaction network} is required to know the extent of this result. 
    \item The cumulative impact of many flares over hundreds of years {appears to} drives a new chemical `steady-state' for certain species in the network, and in other cases, the impact never levels out, {at least over the duration of the models.} Further modeling is required to know the extent of X-ray flare driven chemistry over the disk life time (millions of years), since this model was computationally limited to 500 years. 
    \item In addition, we present a new observationally-motivated X-ray light curve generator, \xgen. \xgen\ is able to generate a stochastic light curve based on an observed energy distribution and flare frequency, along with a user specified flare shape. Given commonalities between stellar flares across the age spectrum, we have made the code publicly available for use in a variety of applications beyond disk chemistry.
\end{enumerate}

{ Among the chemical species that have been detected in protoplanetary disks \citep{mcguire2018}, we find that only HCO$^+$ and N$_2$H$^+$ are likely to be observed in a flare enhanced state (i.e., $\Delta$\fancyN$>1.20$). 
There is a $15.3\%$ and a $1.4\%$ chance to randomly observe HCO$^+$ and N$_2$H$^+$ in a flare-enhanced state, respectively. 
However, these values only consider variability in disk integrated abundance. Some species may be observably variable in local parts of the disk, as discussed in \citet{waggoner2019} on H$_2$O. 

While the observational implications for individual flares may seem low, 
we note that only $\sim$24 species (excluding isotopes) have been detected in protoplanetary disks at this time. Our findings indicate $\gtrsim10\%$ of known species are flare susceptible. As additional species are observed, or unexpected abundance patterns are seen between species, flare effects, especially on gas-phase cations, should be considered. }

\acknowledgments

{We thank the anonymous referees who provided insightful feedback on this paper.}
A.R.W. acknowledges support from the Virginia Space Grant Consortium and the National Science Foundation Graduate Research Fellowship Program under grant No. 1842490. Any opinions,findings, and conclusions or recommendations expressed in this material are those of the author(s)and do not necessarily reflect the views of the National Science Foundation. L.I.C. acknowledges support from the David and Lucille Packard Foundation, Johnson \& Johnson WISTEM2D, and NASA ATP 80NSSC20K0529. We also are grateful for the use of the Hydra Computing Cluster at the University of Virginia, on which the modeling for this work was carried out.

\bibliography{ms}{}
\bibliographystyle{aasjournal}

\appendix

\section{Most variable species}\label{apA}

Table~\ref{tab:my_label} presents the most chemically impacted species in the model. As can be seen, cations are by far the most strongly impacted molecules in the network.

\begin{table}[h!]
    \caption{Left column: The thirty most stochastically variable chemical species in the network. Right column: The thirty species that are most changed when compared to a model without flares after 500 years of chemical evolution.}
    \label{tab:my_label}
    \centering
    \begin{tabular}{lcc|lcc}
    Highest $\sigma$ & & & Highest \change\ & & \\
 & $\sigma$ & \change\ & & $\sigma$ & \change\ \\
 \hline
SO$_2^+$  & 4.51 & 1.00 & H$_2$S$_2^+$ & 0.37 & 1.25 \\
H$_2^+$  & 4.39 & 0.99 & HOCS$^+$ & 0.31 & 1.25 \\
O$_2$H$^+$  & 4.36 & 1.06 & HSO$_2^+$ & 0.37 & 1.24 \\
H$_3^+$  & 3.24 & 0.99 & N$^+$ & 0.42 & 1.22 \\
HeH$^+$  & 1.03 & 1.18 & He$^+$ & 0.13 & 1.21 \\
CO$_2^+$  & 0.70 & 1.01 & HSO$^+$ & 0.23 & 1.21 \\
C$_5$H$_2^+$  & 0.64 & 1.00 & H$^+$ & 0.24 & 1.20 \\
H$_3$O$^+$  & 0.51 & 1.08 & HeH$^+$ & 1.03 & 1.18 \\
N$_2^+$  & 0.46 & 1.10 & COOCH$_4^+$ & 0.23 & 1.17 \\
N$_+$  & 0.42 & 1.22 & HC$_3$O$_+$ & 0.28 & 1.17 \\
HCO$^+$  & 0.40 & 1.07 & CH$_2$O$_2^+$ & 0.05 & 1.17 \\
HNO$^+$  & 0.38 & 1.03 & O$^+$ & 0.20 & 1.16 \\
HSO$_2^+$  & 0.37 & 1.24 & C$_2$H$_5$OH$^+$ & 0.11 & 1.16 \\
H$_2$S$_2^+$  & 0.37 & 1.25 & O$_2$H & 0.07 & 1.13 \\
C$_6$H$_5^+$  & 0.32 & 1.11 & C$_3$H$_3$N$^+$ & 0.05 & 0.87 \\
HOCS$^+$  & 0.31 & 1.25 & HNS$^+$ & 0.22 & 1.13 \\
CH$_4^+$  & 0.28 & 1.12 & CH$_4^+$ & 0.28 & 1.12 \\
HC$_3$O$^+$  & 0.28 & 1.17 & H$_2$CS$^+$ & 0.16 & 1.12 \\
H$^+$  & 0.24 & 1.20 & H$_3$S$_2^+$ & 0.17 & 1.11 \\
HSO$^+$  & 0.23 & 1.21 & C$_6$H$_5^+$ & 0.32 & 1.11 \\
COOCH$_4^+$  & 0.23 & 1.17 & C$_7$H$_4^+$ & 0.02 & 1.10 \\
HNS$^+$  & 0.22 & 1.13 & N$_2^+$ & 0.46 & 1.10 \\
C$_7$H$_2^+$  & 0.22 & 0.99 & C$_6$H$_4^+$ & 0.05 & 1.09 \\
O$^+$  & 0.20 & 1.16 & CH$_3$O$_2^+$ & 0.12 & 1.09 \\
C$_3$H$_4^+$  & 0.19 & 1.02 & C$_2$H$_6^+$ & 0.10 & 1.09 \\
H$_3$S$_2^+$  & 0.17 & 1.11 & C$_4$H$_4$N$^+$ & 0.04 & 0.92 \\
N$_2$H$^+$  & 0.17 & 1.00 & C$_7$H$_5^+$ & 0.02 & 1.08 \\
H$_2$CS$^+$  & 0.16 & 1.12 & C$_4$H$_5^+$ & 0.02 & 1.08 \\
He$^+$  & 0.13 & 1.21 & H$_3$O$^+$ & 0.51 & 1.08 \\
C$_4$H$_7^+$  & 0.12 & 1.06 & HCO$^+$ & 0.40 & 1.07 \\
    \end{tabular}\label{tbl:highestchanges}
\end{table}

\end{document}